\documentclass[nofootinbib,aps,11pt]{revtex4-1}
\usepackage{graphicx}
\usepackage{hyperref}
\usepackage{cancel}
\usepackage{slashed}
\usepackage{amssymb}
\usepackage{textcomp}
\usepackage{amsmath}
\usepackage{bm}
\usepackage{times}
\usepackage{epsfig}
\usepackage{color}
\usepackage{makecell}
\newcommand{\GeV}{{\rm GeV}}
\newcommand{\MeV}{{\rm MeV}}
\newcommand{\TeV}{{\rm TeV}}
\newcommand{\eV}{{\rm eV}}

\newcommand{\fb}{{\rm fb}}
\newcommand{\pb}{{\rm pb}}
\newcommand{\BR}{{\rm BR}}
\newcommand{\Tr}{{\rm Tr}}

\begin{document}
\title{\LARGE Testing Type II Radiative Seesaw Model: from Dark Matter Detection to LHC Signatures}
\bigskip
\author{Shu-Yuan Guo$^{1}$}
\email{shyuanguo@mail.nankai.edu.cn}
\author{Zhi-Long Han$^{1}$}
\email{hanzhilong@mail.nankai.edu.cn}
\author{Yi Liao~$^{1,2,3}$}
\email{liaoy@nankai.edu.cn}

\affiliation{
$^1$~School of Physics, Nankai University, Tianjin 300071, China
\\
$^2$ CAS Key Laboratory of Theoretical Physics, Institute of Theoretical Physics,
Chinese Academy of Sciences, Beijing 100190, China
\\
$^c$ Center for High Energy Physics, Peking University, Beijing 100871, China}
\date{\today}
\begin{abstract}
We analyse the testability of the type II radiative seesaw in which neutrino mass and dark matter (DM) are related at one-loop level. Under the constraints from DM relic density, direct and indirect detection, and invisible Higgs decays, we find three possible regions of DM mass $M_{s_1}$ that can survive the present and even the future experiments: (1) the Higgs resonance region with $M_{s_1}\sim M_h/2$, (2) the Higgs region with $M_{s_1}\sim M_h$, and (3) the coannihilation region with $M_{s_2}\sim M_{s_1}$. Here $s_{1,2}$ are two scalar singlets with the lighter $s_1$ being the DM candidate. Based on DM properties and direct collider constraints, we choose three benchmark points to illustrate the testability of this model at LHC. We perform a detailed simulation of the four-lepton and tri-lepton signatures at $13~(14)~\TeV$ LHC. While both signatures are found to be promising at all benchmark points, the tri-lepton one is even better: it is possible to reach the $5\sigma$ significance with an integrated luminosity of $100~\fb^{-1}$.
\end{abstract}

\maketitle

\section{Introduction}

Tiny but nonzero neutrino masses and nonbaryonic dark matter (DM) provide strong evidence for
physics beyond the standard model (SM). While neutrino masses can be incorporated by a dimension-5 Weinberg operator~\cite{Weinberg:1979sa}, whose tree-level realizations~\cite{Ma:1998dn} correspond to the standard three types of seesaw~\cite{type1,type2,type3}, a DM candidate is missing in these UV completions. On the other hand, weakly interacting massive particles (WIMPs) have long been a leading candidate of DM, due to the coincidence between the observed DM relic density and the thermal abundance of electroweak (EW) scale WIMPs \cite{Jungman:1995df}, which are far below the usual seesaw scales. It would be appealing if neutrino masses and DM are intimately linked and originate at the same EW scale.

A natural pathway to gain neutrino mass at the EW scale is to push it to a radiative effect~\cite{radiative,Boucenna:2014zba}. A specific radiative neutrino mass model with a DM candidate was proposed in Ref.~\cite{Ma:2006km}; see Refs.~\cite{1loop,2loop,3loop,4loop} for more options at the one-, two-, three-, and four-loop level respectively. Usually, a discrete symmetry is imposed by hand so that neutrino mass generation is forbidden at a lower order, as well as to stabilize DM at the same time. The simplest such symmetry is a $\mathbb{Z}_2$ parity, which may also appear as a remnant of a broken local symmetry \cite{Krauss:1988zc}.

In analogy to $R$-parity in supersymmetric (SUSY) models, Ref. \cite{Ma:2015xla} proposed that a dark $\mathbb{Z}_2$ parity, i.e., $(-1)^{L+2j}$, is derivable from lepton parity $(-1)^L$. Here $j$ is the spin and $L$ the lepton number of the particle. Notably, if the radiative generation of neutrino mass is extended to other fermions through a dark matter mediator \cite{Ma:2013mga}, this dark $\mathbb{Z}_2$ parity becomes exactly the $R$-parity. The {\em ad hoc} imposed $\mathbb{Z}_2$ parity in many existing neutrino models with DM \cite{1loop,2loop,3loop}, including radiative versions of type I and III seesaws, is found to correspond to the dark $\mathbb{Z}_2$ parity \cite{Ma:2015xla}. A radiative version of the type II seesaw with DM seems more difficult, because the exact symmetry used to forbid the tree-level coupling $\overline{F_L^C}\xi F_L$, where $F_L$ is the lepton doublet and $\xi$ the scalar triplet, will also prohibit any loop realization of it. The new insight into the relation between dark and lepton parities is that the symmetry used to forbid the hard term $\overline{F_L^C}\xi F_L$ must be softly broken in the loop graphs for neutrino masses~\cite{Ma:2015xla}, so that a radiative realization of the type II seesaw becomes possible. Following this line of reasoning, $\xi$ is assigned with a vanishing lepton number ($L=0$) so that the tree-level coupling $\overline{F_L^C}\xi F_L$ is still forbidden. But the lepton number is broken to $(-1)^L$ by a soft term for the singlet scalars $s_a$ with $L=1$. With the introduction of a fermion doublet $\chi=(N,E)$ with $L=0$, the neutrino mass is indeed generated at one-loop level as shown in Fig. \ref{Fig:mv}. While both $s_a$ and $\chi=(N,E)$ are odd under the dark parity, the lightest scalar singlet $s_1$ is a DM candidate.

The phenomenology of this type II radiative seesaw is quite different from the conventional type II seesaw, and thus deserves a separate study. In particular, it incorporates a DM candidate. In this work we aim to implement a comprehensive analysis on DM properties, including relic density, direct and indirect detection, and invisible Higgs decays. Concerning the LHC observation, the new decay channels of the scalar triplet, e.g., $H^{++}\to E^+E^+$ with $E^+\to\ell^+s_1$, will lead to signatures of multi-lepton plus large missing transverse energy $\cancel{E}_T$. We will perform a detailed simulation at $13~(14)~\TeV$ LHC of the four- and tri-lepton signatures coming from the $H^{++}H^{--}$ pair production and $H^{\pm\pm}H^\mp$ associated production, respectively.

The rest of the paper is organized as follows. In Sec. \ref{SEC:Model}, we review the type II radiative seesaw and discuss constraints from lepton flavor violation (LFV) and direct collider searches. In Sec. \ref{SEC:DM}, we investigate the DM properties under constraints from relic density, direct and indirect detections as well as invisible Higgs decays. In Sec. \ref{SEC:SG}, we study the decay properties of the new particles and then perform a simulation of the four- and tri-lepton signatures at LHC. Finally, our conclusions are presented in Sec.~\ref{SEC:CL}.

\section{Type II Radiative Seesaw}\label{SEC:Model}

\subsection{The Model}

\begin{table}[!htbp]
\begin{tabular}{c c c c c c c}
\hline\hline
\quad Particles\quad & \quad$\Phi$\quad & \quad$F_{iL}$\quad & \quad$\ell_{iR}$\quad & \quad$\xi$\quad & \quad$s_a$\quad & \quad$\chi$\quad
\\\hline
\quad Dark $\mathbb{Z}_2$ & \quad$+$ & \quad$+$ & \quad$+$ & \quad$+$ & \quad$-$ & \quad$-$
\\
\hline
\quad $L$\quad  & \quad$0$ & \quad$1$ & \quad$1$ & \quad$0$ & \quad$1$ & \quad$0$
\\
\hline
\quad $U(1)_Y$\quad  & \quad$1$ & \quad$-1$ & \quad$-2$ & \quad$2$ & \quad$0$ & \quad$-1$
\\
\hline
\quad $SU(2)_L$\quad & \quad$2$ & \quad$2$ & \quad$1$ & \quad$3$ & \quad$1$ & \quad$2$
\\
\hline\hline
\end{tabular}
\caption{Relevant fields and their charge assignments.}
\label{Tab:Content}
\end{table}

\begin{figure}[!htbp]
\begin{center}
\includegraphics[width=0.5\linewidth]{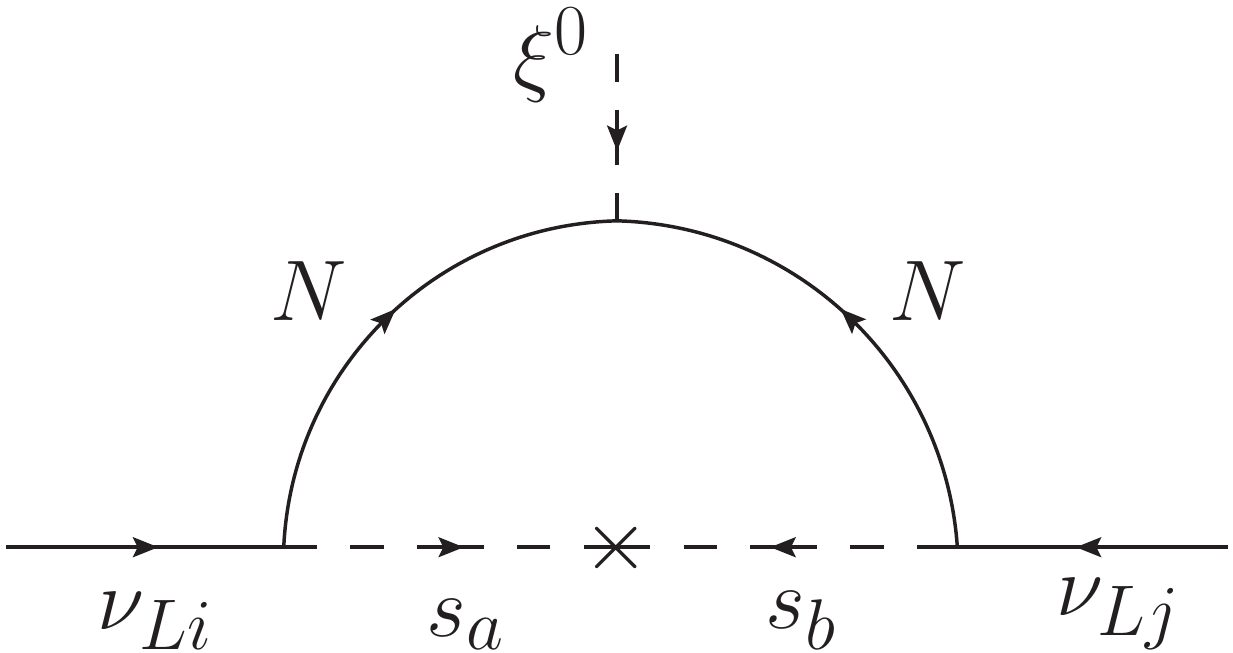}
\end{center}
\caption{Feynman graph for radiative neutrino mass.
\label{Fig:mv}}
\end{figure}

The type II radiative seesaw was proposed in Ref. \cite{Ma:2015xla}, and its phenomenology was briefly discussed in Ref. \cite{Fraser:2015mhb}. In addition to the SM scalar doublet $\Phi$, lepton doublet $F_{iL}$ and singlet $\ell_{iR}$ fields, one scalar triplet $\xi$, two scalar singlets $s_a$, and one vector-like fermion doublet $\chi=(N,E)$ are introduced. The relevant fields and their charge assignments are listed in Table~\ref{Tab:Content}. Differently from the canonical type II seesaw \cite{type2}, the scalar triplet $\xi$ is assigned a vanishing lepton number so that the hard $L$-breaking term $\overline{F_L^C}\xi F_L$ is forbidden at the Lagrangian level and neutrinos remain massless at tree level. But neutrinos can gain a radiative mass with the help of a soft $L$-breaking term $s_a s_b$ and the $L$-conserving couplings $\overline{\chi^C}\xi\chi$ and $s_a\overline{F_L}\chi_R$, as shown in Fig. \ref{Fig:mv}.

The Yukawa couplings and masses for the SM leptons and the new fermion $\chi$ are:
\begin{equation}
\mathcal{L}_{\text{Y}} = - y_{ij}^{\prime}\overline{F_{iL}}\Phi\ell_{jR} - M_{\chi} \bar{\chi}_L\chi_R
 -\frac{1}{2} z^L \overline{\chi_L^C}i\tau^2\xi\chi_L -\frac{1}{2} z^R \overline{\chi_R^C}i\tau^2\xi\chi_R
 -x_{ai}^{\prime} s_a \overline{F_{iL}}\chi_R + \mbox{h.c.}.
\end{equation}
While the charged member $E$ of the doublet $\chi$ has a mass $M_\chi$, its neutral member $N$ mixes by the $z^{L,R}$ couplings into a pair of Majorana particles of generally different masses when $\xi$ develops a vacuum expectation value (VEV), $u$. Since $u\ll M_\chi$, the Majorana particles are almost degenerate with $E$ for all practical purposes. The electroweak precision constraints on $\chi$ are then easily avoided \cite{Bhattacharya:2015qpa}.

In terms of the scalar fields $s_a$ and
\begin{align}
\Phi=\left(
\begin{array}{c}
\phi^+\\
\phi^0
\end{array}\right),\quad \xi =\left(
\begin{array}{cc}
\xi^+/\sqrt{2} & \xi^{++}\\
\xi^0 & -\xi^+/\sqrt{2}
\end{array}\right),
\end{align}
the complete scalar potential invariant under the dark $\mathbb{Z}_2$ is given by
\begin{eqnarray}\label{Eq:potential}\nonumber
V&=& -m^2 \Phi^\dag\Phi + M^2_{\xi}\Tr(\xi^\dag\xi)+(m^2_s)_{ab}s_a^*s_b + (\kappa^{\prime2}_{ab}s_as_b+\mu\Phi^\dag\xi\tilde{\Phi}+\mbox{h.c.})\\
&&+\lambda (\Phi^\dag\Phi)^2+\lambda_1^\xi \left(\Tr(\xi^\dag\xi)\right)^2
  +\lambda_2^\xi \Tr(\xi^\dag\xi)^2+\lambda^{\prime s}_{ab;cd}s_a^*s_b^*s_cs_d
\\\nonumber
&&+\lambda_1^{\Phi\xi}(\Phi^\dag\Phi)\Tr(\xi^\dag\xi)+\lambda_2^{\Phi\xi}\Phi^\dag\xi\xi^\dag\Phi
  + \lambda_{ab}^{\prime s\Phi}s_a^*s_b\Phi^\dag\Phi+\lambda_{ab}^{\prime s\xi}s_a^*s_b\Tr\xi^\dag\xi
\end{eqnarray}
Assuming $m^2$ and $M_\xi^2$ are positive, $\phi^0$ develops a VEV, $\langle\phi^0\rangle=v/\sqrt{2}$, which then induces a VEV for $\xi^0$, $\langle\xi^0\rangle=u/\sqrt{2}$, through the $\mu$ term with $u=\mu v^2 /(\sqrt{2}M_{\xi}^2)$. We further assume $m_s^2$, $\kappa'$, $\lambda^{\prime s\Phi}$, $\lambda^{\prime s\xi}$ are such that $s_a$ will not develop a VEV to avoid spontaneous breaking of the lepton number $L$~\cite{SVL}. In contrast to the conventional seesaw \cite{type2}, the $\mu$ term does not break $L$ so that in principle it is not necessarily small. But since $u$ is constrained by the $\rho$ parameter to be small, $u\leq5~\GeV$ \cite{Agashe:2014kda}, the easiest way to accomplish this is still to assume a small $\mu$.

The masses of the SM Higgs boson $h$ and the scalar triplet $\xi$ are hardly affected by a small $u$, while the spectra of $\xi$ depend on $v$ through the couplings $\lambda_{1,2}^{\Phi\xi}$. In the following study, we will be interested in a relatively heavy scalar triplet with $M_{\xi}^2>v^2/2$. For simplicity, we ignore the contributions from $\lambda_{1,2}^\xi$ and $\lambda_{1,2}^{\Phi\xi}$, so that all members of $\xi$ are approximately degenerate, easily fulfilling the electroweak precision constraints~\cite{Kanemura:2012rs,Chun:2012jw,Aoki:2012jj}. We refer to Refs.~\cite{Arhrib:2011uy,Bonilla:2015eha,Chabab:2015nel,Haba:2016zbu,Das:2016bir} for a detailed study of the scalar potential and Refs.~\cite{Han:2015hba,Akeroyd:2011zza,Melfo:2011nx,
Aoki:2011pz,Akeroyd:2012nd,Chun:2012zu} for phenomenology of a nondegenerate triplet $\xi$ in the type II seesaw model.

After the electroweak symmetry breaking, $\Phi$ and $\xi$ mix into physical scalars ($h,~H^0,~A^0,~H^\pm$) and would-be Goldstone bosons ($G^{0,\pm}$) as:
\begin{align}
&\left(\!\!
\begin{array}{c}
\phi^{\pm}\\
\xi^{\pm}
\end{array}\!\!\right)=R(\theta_+)\left(\!\!
\begin{array}{c}
G^{\pm}\\
H^{\pm}
\end{array}\!\!\right),~
\sqrt{2}\left(\!\!
\begin{array}{c}
\textrm{Im }\phi^0\\
\textrm{Im }\xi^0
\end{array}\!\!\right)=R(\alpha)\left(\!\!
\begin{array}{c}
G^{0}\\
A^{0}
\end{array}\!\!\right),~
\sqrt{2}\left(\!\!
\begin{array}{c}
\textrm{Re }\phi^0\\
\textrm{Re }\xi^0
\end{array}\!\!\right)=R(\theta_0)\left(\!\!
\begin{array}{c}
h\\
H^0
\end{array}\!\!\right),
\end{align}
where $R(\theta_+)$, $R(\alpha)$, and $R(\theta_0)$ are rotation matrices with the corresponding mixing angles given by,
\begin{align}
\tan \theta_+ = \frac{\sqrt{2} u}{v},~
\tan \alpha = \frac{2 u}{v},~
\tan 2\theta_0 \approx\frac{u}{v} \frac{4M_{\xi}^2}
{M_{\xi}^2-M_{h}^2}.
\end{align}
Here, $h$ is regarded as the boson discovered at LHC \cite{Aad:2012tfa,Chatrchyan:2012xdj} with mass $M_h=125~\GeV$ \cite{Aad:2015zhl}. The physical particles also include the doubly-charged scalars $H^{\pm\pm}\equiv\xi^{\pm\pm}$. Due to the dark $\mathbb{Z}_2$ symmetry, $s_a$ do not mix with $\Phi$ and $\xi$. Considering $u\ll v$, the Hermitian mass squared matrix of $s_a$ is given by
\begin{equation}
(M_s^2)_{ab}=(m_s^2)_{ab}+\frac{1}{2}\lambda^{\prime s\Phi}_{ab}v^2,
\end{equation}
while the supposedly small $\kappa^{\prime 2}$ term contributes to the mass splitting between the real and imaginary parts of $s_a$. When the matrix $M_s^2$ is diagonalized to its eigenvalues $M_{s_1}^2,M_{s_2}^2$, the coupling matrix $\lambda^{\prime s\Phi}$ is not diagonal in general. As to be shown in Sec.~\ref{SEC:DM} , the off-diagonal coupling will play an important role in dark matter phenomenology. From now on we will remove the prime from couplings associated with diagonaized fields.

The one-loop induced neutrino masses shown in Fig. \ref{Fig:mv} are calculated as \cite{Ma:2015xla,Fraser:2015mhb}
\begin{equation}\label{mv}
m_\nu = \frac{u x^2\kappa^2}{8\sqrt{2}\pi^2 m_\chi^2}\left[z^L F_L\left(\frac{m_s^2}{m_\chi^2}\right)+z^R F_R\left(\frac{m_s^2}{m_\chi^2}\right)\right],
\end{equation}
where the loop functions $F_L$ and $F_R$ are given by
\begin{eqnarray}
F_L(x)&=&\frac{1}{(1-x)^3}\big[2(1-x)+(1+x)\ln x\big],
\\
F_R(x)&=&\frac{1}{(1-x)^3}\big[1-x^2+2x\ln x\big].
\end{eqnarray}
In order to obtain $m_\nu\sim0.01~\eV$, we can take, for instance, $u\sim0.5~\GeV$, $x\sim0.005$, $\kappa\sim3~\GeV$, and $z^{L,R}\sim1$ with both $M_s$ and $M_\chi$ around the EW scale.

\subsection{Constraints}

The $\overline{F_L^C}\xi F_L$ coupling responsible for LFV processes in the type II seesaw \cite{Chun:2003ej} is missing in the current radiative seesaw. Instead, the LFV transitions of charged leptons are now mediated by charged fermions $E^\pm$ and singlet scalars $s_a$ through the Yukawa coupling $x$. For instance, the branching ratio of the lepton radiative decay $\ell_j\to\ell_i\gamma$ is calculated as \cite{Ding:2014nga}:
\begin{equation}
\mbox{BR}(\ell_j\to \ell_i\gamma)=\mbox{BR}(\ell_j\to \ell_i \bar{\nu}_i\nu_j)
\frac{3\alpha}{16\pi G_F^2M_\chi^4}
\left|\sum_{a}x_{ai}^{*}x_{aj}F\left(\frac{M_{s_a}^2}{M_\chi^2}\right)\right|^2,
\end{equation}
with the loop function $F(x)$ given by:
\begin{equation}
F(x)=-\frac{1}{12(1-x)^4}[1-6x+3x^2+2x^3-6x^2\ln x].
\end{equation}
The most stringent limit comes from $\mu\to e\gamma$ with the upper bound $\BR(\mu\to e\gamma)<4.2\times10^{-13}$~\cite{MEG}, which in turn requires, for an order of magnitude estimate, that
\begin{equation}
\Big|x_{ae}^*x_{a\mu}\Big|\lesssim 5\times10^{-5} \left(\frac{M_\chi}{100~\GeV}\right)^2,
\end{equation}
for $M_{s_a}\sim M_{\chi}$. Therefore, when $M_{\chi}\sim 200~\GeV$, the Yukawa coupling is restricted to $|x_{ai}|\lesssim0.01$ without requiring a special flavor structure. For simplicity, we will assume a universal Yukawa coupling $x_{ai}=0.005$ in the following discussion. In particular, our benchmark points fully satisfy this constraint.

A distinct feature of the type II seesaw is the presence of doubly-charged scalars $H^{\pm\pm}$, which has been extensively studied by theory \cite{Hpp:ph} and experiment \cite{ATLAS:2012hi,Chatrchyan:2012ya} groups. The most promising decay channel of $H^{\pm\pm}$ is the same-sign dilepton channel $H^{\pm\pm}\!\!\to\ell^\pm\ell^\pm$. Based on this channel, a lower bound on the mass of $H^{\pm\pm}$ is set by ATLAS~\cite{ATLAS:2012hi} to be about $490$ to $550~\GeV$ assuming $100\%$ decays into $e^\pm e^\pm,~e^\pm\mu^\pm,~\mu^\pm\mu^\pm$, and is extended to between $608$ and $621~\GeV$ by CMS \cite{Chatrchyan:2012ya}. In the type II radiative seesaw, the same-sign diboson channel $H^{++}\!\!\to W^+ W^+$ is dominant when $M_{\xi}<2M_\chi$, since $H^{\pm\pm}\!\!\to \ell^\pm\ell^\pm$ is one-loop suppressed. In this case, the lower bound on $M_{H^{++}}$ derived from the same-sign dilepton channel is much weaker, about $84-90~\GeV$ \cite{Kanemura:2013vxa}.

\begin{figure}[!htbp]
\begin{center}
\includegraphics[width=0.5\linewidth]{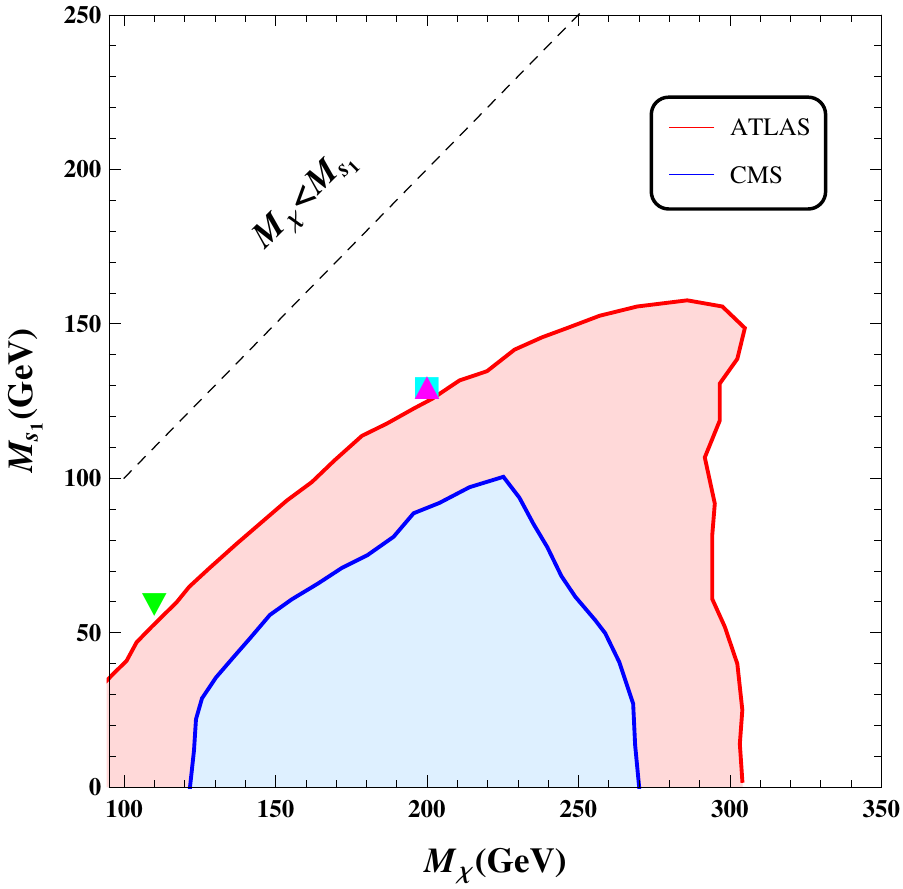}
\end{center}
\caption{Exclusion regions in the $M_{s_1}\!\!\!-\!M_\chi$ plane by ATLAS and CMS. The benchmark points in Table \ref{Tab:BP} are also indicated.
\label{Dilepton}}
\end{figure}

When $M_{\xi}>2M_{\chi}$, the new decay channel $H^{++}\!\!\to E^+E^+$ with the subsequent decay $E^+\!\to \ell^+ s$ will be dominant, resulting in the signature of a same-sign dilepton plus missing transverse energy $\ell^+\ell^++\cancel{E}_T$. The direct pair production of $E^\pm$ leads to the collider signature $pp\to E^+E^-\to \ell^+\ell^-+\cancel{E}_T$, which in SUSY models could arise from slepton ($\tilde{\ell}$) pair production followed by decays $\tilde{\ell}^\pm\to\ell^{\pm}\tilde{\chi}_1^0$, where the lightest neutrolino $\tilde{\chi}_1^0$ appears as missing transverse energy. In Fig. \ref{Dilepton}, we show the excluded regions in the $M_{s_1}\!\!-\!M_{\chi}$ plane by ATLAS \cite{Aad:2014vma} and CMS \cite{Khachatryan:2014qwa} based on simplified SUSY models. Assuming exclusive decays into $e$ or $\mu$, CMS has excluded $120~\GeV\lesssim M_{\chi}\lesssim 260~\GeV$ for $M_{s_1}\lesssim 50~\GeV$, while for $M_{s_1}\gtrsim 100~\GeV$, no limit on $M_{\chi}$ has been available~\cite{Khachatryan:2014qwa}. Compared to CMS, ATLAS has set a more stringent bound, i.e., $M_{\chi}\lesssim 300~\GeV$ with a light $s_1$ is excluded \cite{Aad:2014vma}. Nevertheless, a compressed spectrum with $M_{\chi}\sim M_{s_1}$ is still allowed for $M_{\chi}\sim 200~\GeV$. Considering this, we choose three benchmark points shown in Table \ref{Tab:BP} and Fig. \ref{Dilepton} for the study of dark matter in Sec. \ref{SEC:DM} and of collider signatures in Sec. \ref{SEC:SG}.

\begin{table}[!htbp]
\begin{tabular}{|c|c|c|c|c|c|c|c|c|c|}
\hline
  & $M_{s_1}$ & $M_{s_2}$ & $M_{\chi}$ & $M_\xi$ & $\lambda_{11}^{s\Phi}$ & $\lambda_{12}^{s\Phi}$ & $\Omega_{\text{DM}} h^2$ & $\sigma^{\textrm{SI}}$ & Marker \\
\hline
BP-A& 60 & 200 & 110 & 400 &0.00095 & 0.05 & 0.1177 & $2.1$ &{\color{green}$\blacktriangledown$} \\
\hline
BP-B& 130 & 250 & 200 & 410 & 0.010 & 0.34 & 0.1186 & $51$
&{\color{magenta}$\blacktriangle$}\\
\hline
BP-C& 130 & 300 & 200 & 500 & 0.005 & 0.40 & 0.1172 & $13$
&{\color{cyan}$\blacksquare$} \\
\hline
\end{tabular}
\caption{Benchmark points for the study of dark matter and collider signatures. All masses are in units of GeV and $\sigma^{\textrm{SI}}$ in $10^{-12}~\textrm{pb}$.}
\label{Tab:BP}
\end{table}

\section{Dark Matter}\label{SEC:DM}

The lightest inert scalar $s_1$ is a DM candidate in the type II radiative seesaw. To investigate the DM phenomenology, we implement the model into {\tt FeynRules} \cite{Christensen:2008py} with the output of a {\tt CalcHEP} \cite{Belyaev:2012qa} model file taken by {\tt micrOMEGAs} \cite{Belanger:2014vza} to evaluate DM variables. Before we move on to scan the parameter space, we give a brief overview of the annihilation channels, which can be classified into five categories:
\begin{itemize}
  \item $s_1^* s_1\to \ell^+\ell^-,\nu \nu$ mediated by the inert fermion doublet $\chi=(N,E)$ via the Yukawa coupling $x_{ai}$. To acquire the correct relic density, a hierarchical structure $|x_{ae}|\lesssim|x_{a\mu}|\lesssim|x_{a\tau}|$ with $|x_{a\tau}|\sim\mathcal{O}(1)$ should be satisfied under the tight constraint from LFV \cite{Vicente:2014wga} if this category is dominant. With our simple assumption of a universal Yukawa coupling $x_{ai}=0.005$,  the contribution to the relic density is safely negligible.
  \item $s_1^* s_1 \to H^{++}H^{--}, H^{+}H^{-}, H^{0}H^{0}, A^{0}A^{0}$ through contact interactions via the quartic coupling $\lambda^{s\xi}$. For our interested decay channel $H^{++}\to E^+E^+$ with $E^+\to \ell^+ s_1$ at LHC, these annihilation channels are kinematically closed. In the scanning of the parameter space, we simply set $\lambda^{s\xi}=0$, thus ignoring this category technically.
  \item $s_1^*s_1\to W^+W^-,b\bar{b}, ...,$ SM pairs mediated by the $s-$channel SM Higgs $h$ via the quartic coupling $\lambda^{s\Phi}_{11}$. For the $s_1^*s_1\to hh$ channel, there is also a contribution from the $t-$channel $s_1$ exchange as well as from the contact term $\sim\lambda_{11}^{s\Phi}s_1^*s_1 hh$. This category corresponds to the well studied Higgs-portal singlet scalar DM~\cite{Silveira:1985rk}. A recent fitting of experimental limits shows~\cite{Cline:2013gha} that the allowed mass region of $M_{s_1}$ is either near $M_h/2\sim 53-62.5~\GeV$ or larger than $185~\GeV$ under the LUX2013 limit. Furthermore, the high mass region between $185~\GeV$ and $3~\TeV$ could be excluded by the forthcoming XENON1T~\cite{Aprile:2012zx}, and the allowed low mass region could be shrunk to $\sim 55-62.5~\GeV$~\cite{Cline:2013gha}.
  \item $s_1^*s_1\to hh$ mediated by the heavier inert scalar $s_2$ in the $t-$channel via the quartic coupling $\lambda^{s\Phi}_{12}$. An amazing feature of this category is that it offers a new annihilation channel without affecting the DM-nucleon cross section~\cite{Fraser:2015mhb}. Therefore, when this category is dominates the relic density, the $s_1$ DM can escape the stringent constraints from direct detection and rescue the high mass region $M_{s_1}>M_h$.
  \item $s_1^* s_2,s_2^* s_1\to W^+W^-, b\bar{b},...,$ SM pairs mediated by the $s$-channel SM Higgs $h$ also via the quartic coupling $\lambda^{s\Phi}_{12}$ when $M_{s_2}\approx M_{s_1}$. This category is the so-called coannihilation \cite{Griest:1990kh}, which can play a crucial role in the relic density. As will be shown later, this category could escape both constraints from direct and indirect detection of DM.
\end{itemize}

In summary, for the consideration of DM phenomenology, we fix the following input parameters \begin{eqnarray}
x_{ai}=0.005,~M_\xi=500~\GeV,~\lambda^{s\xi}=0,~\lambda^{s\Phi}_{22}=0.01,
\end{eqnarray}
and vary other parameters as
\begin{eqnarray}\nonumber
&&M_{s_1}\in[10,1000]~\GeV,~M_{s_2}\!\!-M_{s_1}\in[1,1000]~\GeV,
\\
&&\lambda^{s\Phi}_{11}\in[0.001,1],~\lambda^{s\Phi}_{12}\in[0.001,1].
\end{eqnarray}
We randomly scan over the above parameter space, and impose constraints from relic density, direct detection, indirect detection as well as invisible Higgs decays. We require that the relic density satisfies the combined Planck+WP+highL+BAO result in $2\sigma$ range, i.e., $0.1153<\Omega_{\mbox{\tiny DM}} h^2<0.1221$~\cite{Ade:2013zuv}. For the direct detection, we consider the most restrictive spin-independent limits provided by LUX~\cite{Akerib:2013tjd,Akerib:2016vxi} at present and XENON1T~\cite{Aprile:2012zx} in the future. Meanwhile, the indirect detection limits are taken from Fermi-LAT~\cite{Ackermann:2015zua,Ackermann:2015lka} and HESS~\cite{Abramowski:2013ax}, and the limits on invisible Higgs decay are from the fitting results of visible Higgs decays~\cite{Khachatryan:2014jba}.

\subsection{Direct Detection}
\begin{figure}[!htbp]
\begin{center}
\includegraphics[width=0.5\linewidth]{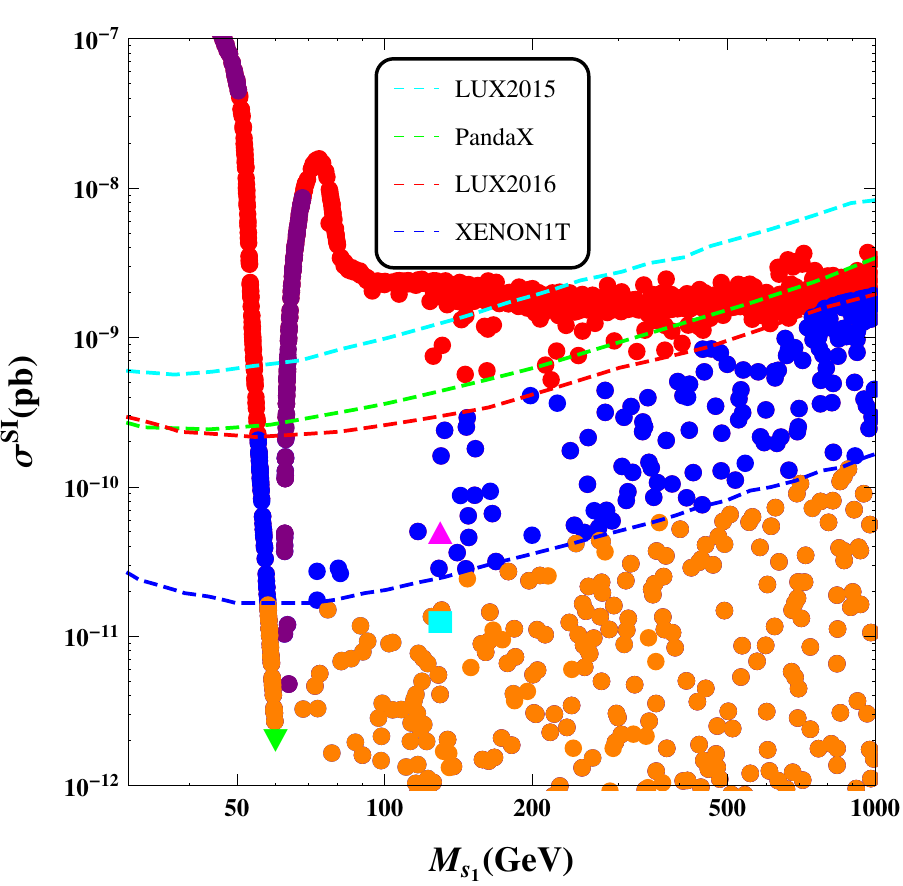}
\end{center}
\caption{Scanned results shown for $\sigma^{\text{SI}}$. The cyan, green, red, and blue lines correspond to LUX2015~\cite{Akerib:2013tjd}, PandaX-II~\cite{Tan:2016zwf}, LUX2016~\cite{Akerib:2016vxi}, and XENON1T~\cite{Aprile:2012zx} limits, respectively. The purple points are excluded by Fermi-LAT~\cite{Ackermann:2015zua}. The predictions at the three benchmark points in Table~\ref{Tab:BP} are also indicated.
\label{DM_DD}}
\end{figure}
The cross section for spin-independent scattering of a scalar singlet $s_1$ on the nucleon is given by \cite{Cline:2013gha}
\begin{equation}
\sigma^{\text{SI}}=\frac{(\lambda^{s\Phi}_{11})^2 f_N^2 \mu_N^2 m_N^2}{\pi M_h^2 M_{s_1}^2},
\end{equation}
where $f_N=0.3$ is the nucleon matrix element, $\mu_N=m_N M_{s_1}/(m_N+M_{s_1})$ the reduced mass, and $m_N=(m_p+m_n)/2=939~\MeV$ the average nucleon mass. In Fig. \ref{DM_DD}, we show the scanning result of $\sigma^{\mbox{\tiny SI}}$ together with the bounds from LUX2015 \cite{Akerib:2013tjd}, LUX2016 \cite{Akerib:2016vxi}, PandaX-II \cite{Tan:2016zwf}, and XENON1T \cite{Aprile:2012zx}. The red and blue points are those that are successively excluded by the current LUX2016 and expected XENON1T limits respectively, while the orange points survive both direct detections and the indirect detection by Fermi-LAT. The upper edge of the distribution corresponds to the minimal Higgs-portal singlet scalar DM \cite{Silveira:1985rk}, with the only two variables being $\lambda^{s\Phi}_{11}$ and $M_{s_1}$. It is clear that the existence of $s_2$ could make the predicted value of $\sigma^{\textrm{SI}}$ much smaller than this minimal case. Considering the current most restrictive limit from LUX2016~\cite{Akerib:2016vxi}, the low mass region $M_{s_1}\lesssim55~\GeV$ and high mass region $64~\GeV\lesssim M_{s_1}$ with $\sigma^{\text{SI}}\gtrsim2.2\times10^{-10}~\pb$ have been excluded. Note that in the minimal singlet scalar DM \cite{Silveira:1985rk}, the high mass region below $1~\TeV$ is now fully excluded by LUX2016. With the future XENON1T limit down to about $10^{-11}~\pb$, the allowed Higgs resonance region will be further narrowed to $58~\GeV\lesssim M_{s_1} \lesssim 62.5~\GeV$. But in this type II radiative seesaw, the $t$-channel exchange of $s_2$ as well as the coannihilation of $s_1$ and $s_2$ could save the high mass region above $M_{h}/2$ to some extent.

In Fig. \ref{DM_DD}, the predictions for $\sigma^{\text{SI}}$ at the three benchmark points in Table~\ref{Tab:BP} are also indicated. These points are representative of three different regions of interest: (1) BP-A stands for the undetectable Higgs resonance region in direct detection experiments, (2) BP-B is for the region that escapes the LUX2016 limit but is within the reach of XENON1T in the high mass region, and (3) BP-C is for one that is even beyond the reach of XENON1T in the high mass region. Note that the masses of $M_{s_1}$ for BP-B and BP-C in the minimal singlet scalar DM model have already been excluded by the LUX2016 experiment. As will be shown in Sec.~\ref{SEC:SG}, the three benchmark points are quite promising to probe at LHC with multi-lepton signatures.

\begin{figure}[!htbp]
\begin{center}
\includegraphics[width=0.33\linewidth]{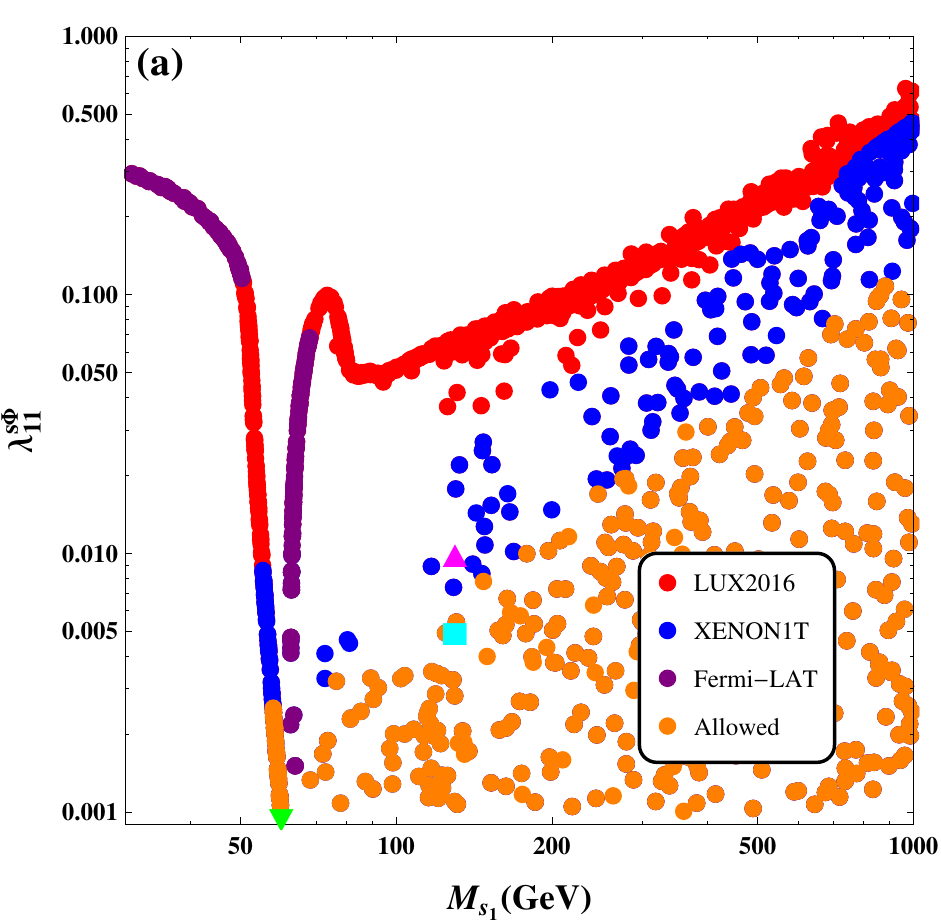}
\includegraphics[width=0.33\linewidth]{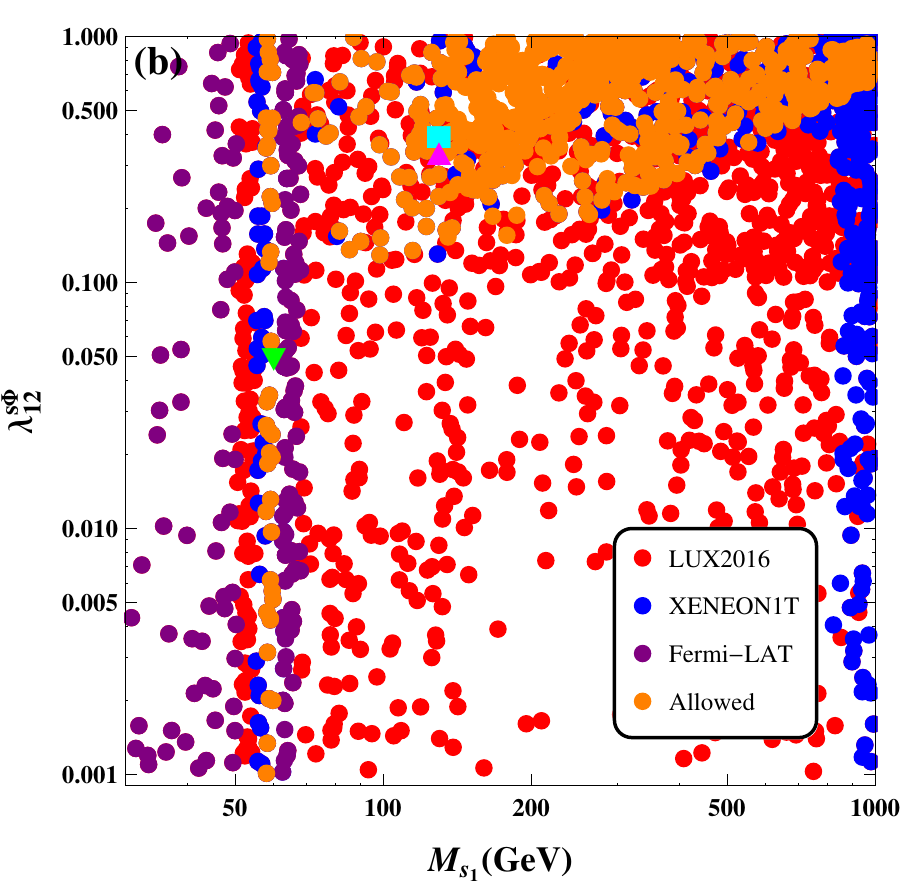}
\includegraphics[width=0.33\linewidth]{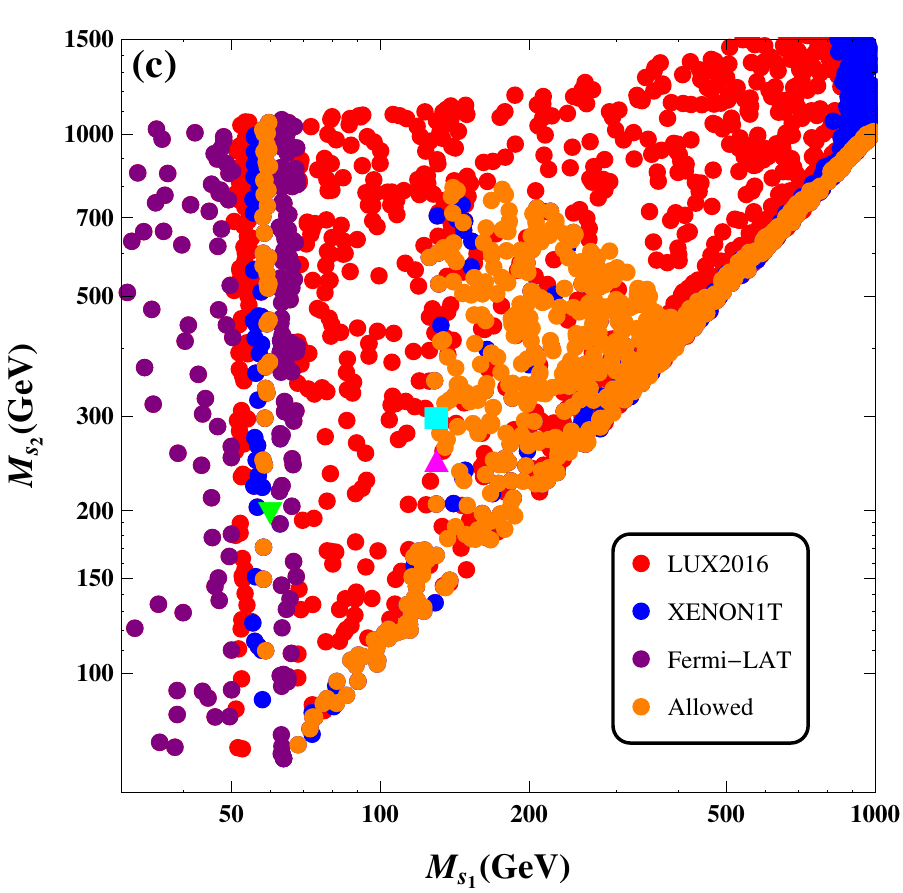}
\end{center}
\caption{Scanned results shown in the plane of $\lambda^{s\Phi}_{11}\!-\!M_{s_1}$ (a), $\lambda^{s\Phi}_{12}\!-\!M_{s_1}$ (b), and $M_{s_2}\!\!-\!M_{s_1}$ (c). The red and blue points are excluded successively by LUX2016~\cite{Akerib:2016vxi} and XENON1T~\cite{Aprile:2012zx}, and the purple points excluded by Fermi-LAT~\cite{Ackermann:2015zua}, while the orange points are still allowed.
\label{DM_Para}}
\end{figure}

In Fig. \ref{DM_Para}, the distributions of our sampled results are depicted in the plane of $\lambda^{s\Phi}_{11}\!-\!M_{s_1}$ (a), $\lambda^{s\Phi}_{12}\!-\!M_{s_1}$ (b), and $M_{s_2}\!\!-\!M_{s_1}$ (c), respectively. In the Higgs resonance region, LUX2016 has excluded $\lambda^{s\Phi}_{11}\gtrsim0.01$, and XENON1T will push this limit down to about $\lambda^{s\Phi}_{11}\gtrsim0.003$. Meanwhile, $\lambda^{s\Phi}_{12}$ and $M_{s_2}$ are free to choose, since $s_2$ does not contribute to the annihilation of $s_1$ in this low mass region. In the high mass region, LUX2016 has excluded some area in the $\lambda^{s\Phi}_{11}\!-\!M_{s_1}$ plane, e.g., $\lambda^{s\Phi}_{11}\gtrsim0.05$ for $M_{s_1}\sim200~\GeV$ and  $\lambda^{s\Phi}_{11}\gtrsim0.5$ for $M_{s_1}\sim1~\TeV$. And the expected XENON1T exclusion limit will be $4\sim5$ times tighter than the current LUX2016 limit. As clearly shown in Fig. \ref{DM_Para} (b), the high mass region $64~\GeV\lesssim M_{s_1}\lesssim850~\GeV$ with $\lambda^{s\Phi}_{12}\lesssim0.15$ has been excluded by LUX2016. And for those that pass the XENON1T limit, we find that the larger $M_{s_1}$ is, the higher the lower limit on $\lambda^{s\Phi}_{12}$ is in the high mass region. From the tight XENON1T constraints on quartic couplings, e.g., $\lambda_{11}^{s\Phi}\lesssim0.01$ and $\lambda_{12}^{s\Phi}\gtrsim0.15$ at $M_{s_1}\sim 200~\GeV$, the dominant annihilation categories for the allowed points are expected to be the $t$-channel $s_2$ exchange and coannihilation channels. In the $M_{s_2}\!\!-\!M_{s_1}$ plane shown in Fig.~\ref{DM_Para}~(c), we find that the allowed points are confined in a triangle area defined by $M_{s_1}\gtrsim M_h$, $M_{s_2}\gtrsim M_{s_1}$, and $M_{s_1}+M_{s_2}\lesssim850~\GeV$ besides the coannihilation area with $M_{s_2}\sim M_{s_1}$, thus the only possible category for this triangle region is the $t$-channel $s_2$ exchange. The upper edge of the triangle corresponds to $\lambda^{s\Phi}_{12}=1$, and $M_{s_1}$ should be less than about $400~\GeV$ in this triangle area. Of course a larger than one value of $\lambda^{s\Phi}_{12}$ or introduction of a third heavy singlet scalar $s_3$ could extend this triangle area. On the other hand, the coannihilation-dominated area with $M_{s_2}\sim M_{s_1}$ can always escape direct detection constraints as shown clearly in Fig.~\ref{DM_Para}~(c).

\subsection{Indirect Detection}

\begin{figure}[!htbp]
\begin{center}
\includegraphics[width=0.45\linewidth]{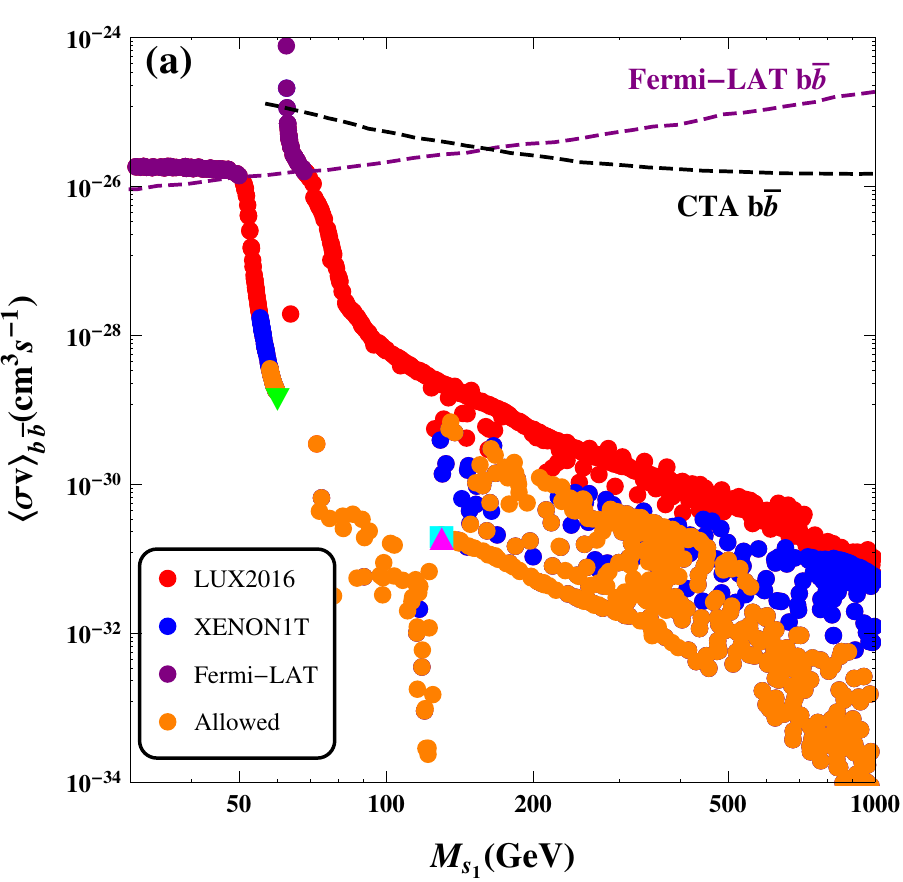}
\includegraphics[width=0.45\linewidth]{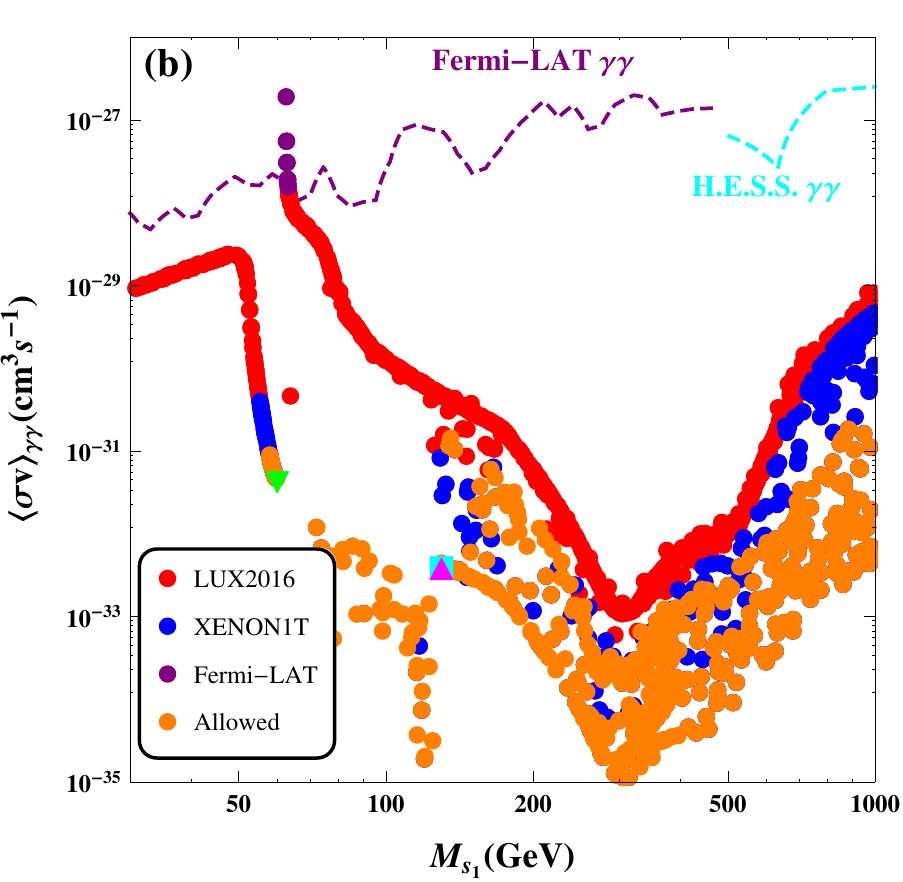}
\includegraphics[width=0.45\linewidth]{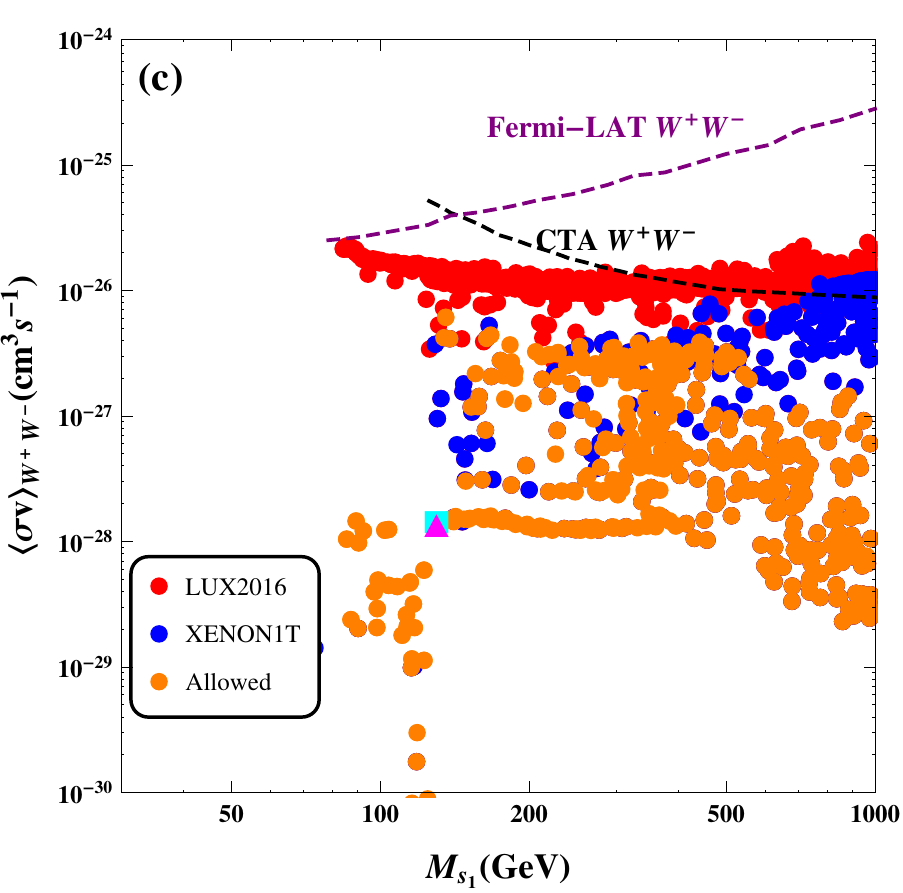}
\includegraphics[width=0.45\linewidth]{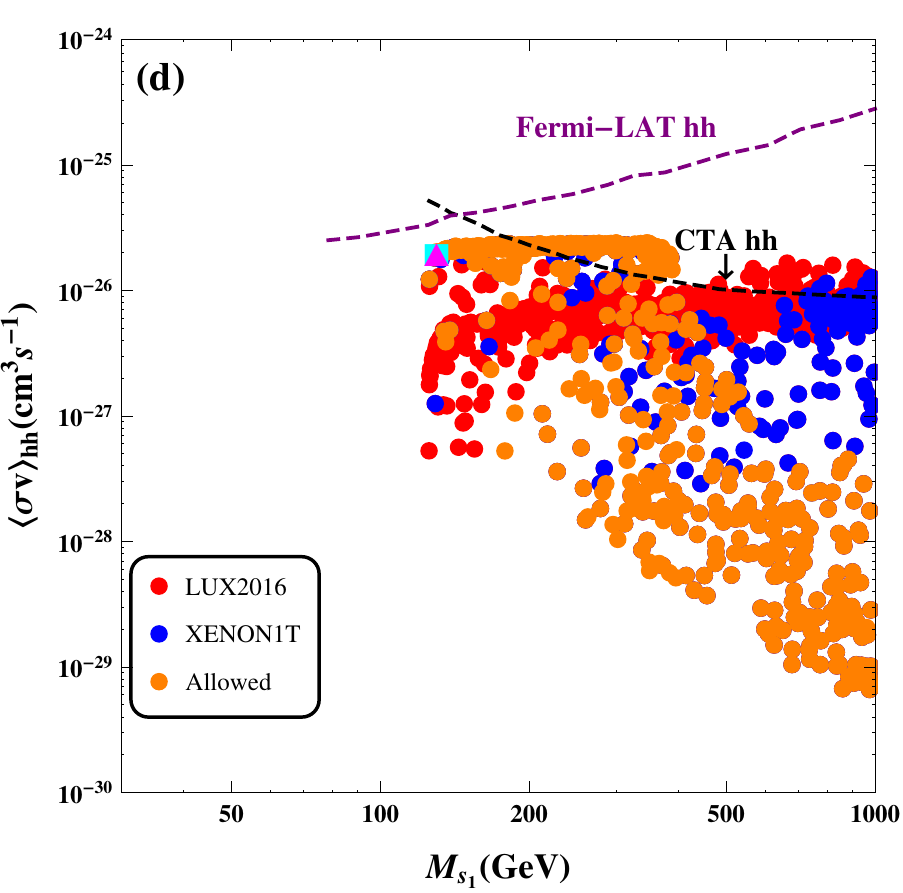}
\end{center}
\caption{Scanned results shown for velocity-averaged annihilation cross section times velocity $\langle\sigma v\rangle$ into $b\bar{b}$ (a), $\gamma\gamma$ (b), $W^+W^-$ (c), and $hh$ (d), using the same legends as in Fig.~\ref{DM_Para}. The dashed curves are upper bounds from Fermi-LAT~\cite{Ackermann:2015zua,Ackermann:2015lka}, HEES~\cite{Abramowski:2013ax}, and CTA~\cite{Doro:2012xx}. The bound on $\langle\sigma v\rangle_{hh}$ is estimated by assuming a similar $\gamma$-spectrum in the $hh$ channel as in the $W^+W^-$ channel~\cite{Cirelli:2010xx}.
\label{DM_Sigmav}}
\end{figure}

The annihilation of the scalar singlet DM $s_1$ into pairs of SM particles also offers an opportunity for indirect detection. In Fig. \ref{DM_Sigmav}, we show the model predictions for $\langle\sigma v\rangle$ in the annihilation channels of $b\bar{b},~\gamma\gamma,~W^+W^-,~hh$ and the corresponding bounds from the Fermi-LAT~\cite{Ackermann:2015zua,Ackermann:2015lka} and HEES~\cite{Abramowski:2013ax} collaborations. The proposed Cherenkov Telescope Array (CTA) experiment \cite{Doro:2012xx} is also included with its most optimistic limits to illustrate future indirect detection potential. As pointed out in previous work \cite{Cline:2013gha}, the indirect constraints are important to exclude the Higgs resonance region where $M_{s_1}\gtrsim M_h/2$.

As clearly shown in Fig. \ref{DM_Sigmav}, the current constraints on $M_{s_1}$ from $\gamma\gamma$, $W^+W^-$, and $hh$  channels are less strict than the $b\bar{b}$ channel, so we first focus on the latter. The Fermi-LAT bound on $\langle\sigma v\rangle_{b\bar{b}}$ has excluded the region $M_{s_1}<50~\GeV$ and $M_h/2\lesssim M_{s_1}<68~\GeV$; see the purple points in Fig.~\ref{DM_Sigmav}(a). Actually, for $M_{s_1}<50~\GeV$, it has already been excluded by LUX2016 (see Fig.~\ref{DM_DD}) as well as invisible Higgs decays (see Fig.~\ref{DM_inv}). For the high mass region above $M_h$, Fermi-LAT can hardly set any limit, since the dominant (co)annihilation final states will be $W^+W^-$ and $hh$ in the type II radiative seesaw. From Fig.~\ref{DM_Sigmav}(c), we see that the CTA limit on $\langle\sigma v\rangle_{W^+W^-}$ is less stringent than the current LUX2016 limit for $M_{s_1}\lesssim 700~\GeV$, and less than the expected XENON1T limit below $1~\TeV$. But for the $hh$ final state, CTA has the potential to further exclude $M_{s_1}\gtrsim180~\GeV$ when $s_1s_1^*\to hh$ mediated by the $t$-channel exchange of $s_2$ is totally dominant at those points that are still allowed by XENON1T. On the other hand, the coannihilation region is always safe to escape the indirect detection. From Fig.~\ref{DM_Sigmav}, one also sees that the three benchmark points in Table \ref{Tab:BP} are on the safe side of indirect detections.

A gamma-ray excess from the galactic center (GCE) was reported by some theoretical analyses \cite{Daylan:2014rsa} and has been recently confirmed by the Fermi collaboration~\cite{TheFermi-LAT:2015kwa}. Although there are various astrophysical explanations to the excess~\cite{GCE-astro}, it is natural to ask if it could be accommodated by DM annihilation~\cite{Hooper:2011ti}. In the type II radiative seesaw under consideration, $s_1$ might play such a role with $M_{s_1}\approx M_h/2$~\cite{Cuoco:2016jqt}. But as a matter of fact, the GCE spectrum is best fit by the $b\bar{b}$ final state for a DM mass of $30-50~\GeV$ with $\langle \sigma v \rangle_{b\bar{b}}\in[1.4,2]\times 10^{-26}\textrm{cm}^3 \text{s}^{-1}$~\cite{Daylan:2014rsa}, which has unfortunately been excluded by Fermi-LAT, LUX, and invisible Higgs decays. A possible solution might be to add a light scalar $\varphi$ with $L=0$ as a mediator \cite{Balazs:2014jla}, which could help $s_1$ avoid conflicts with the current experimental bounds.

\subsection{Invisible Higgs Decays}

\begin{figure}[!htbp]
\begin{center}
\includegraphics[width=0.45\linewidth]{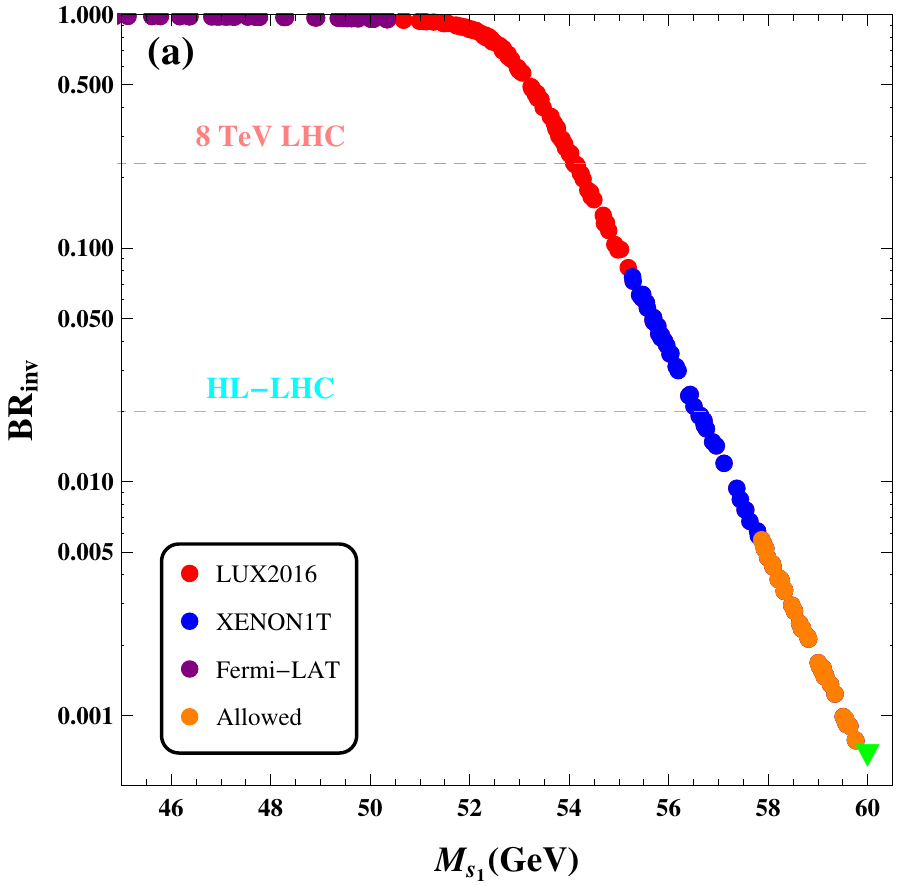}
\includegraphics[width=0.455\linewidth]{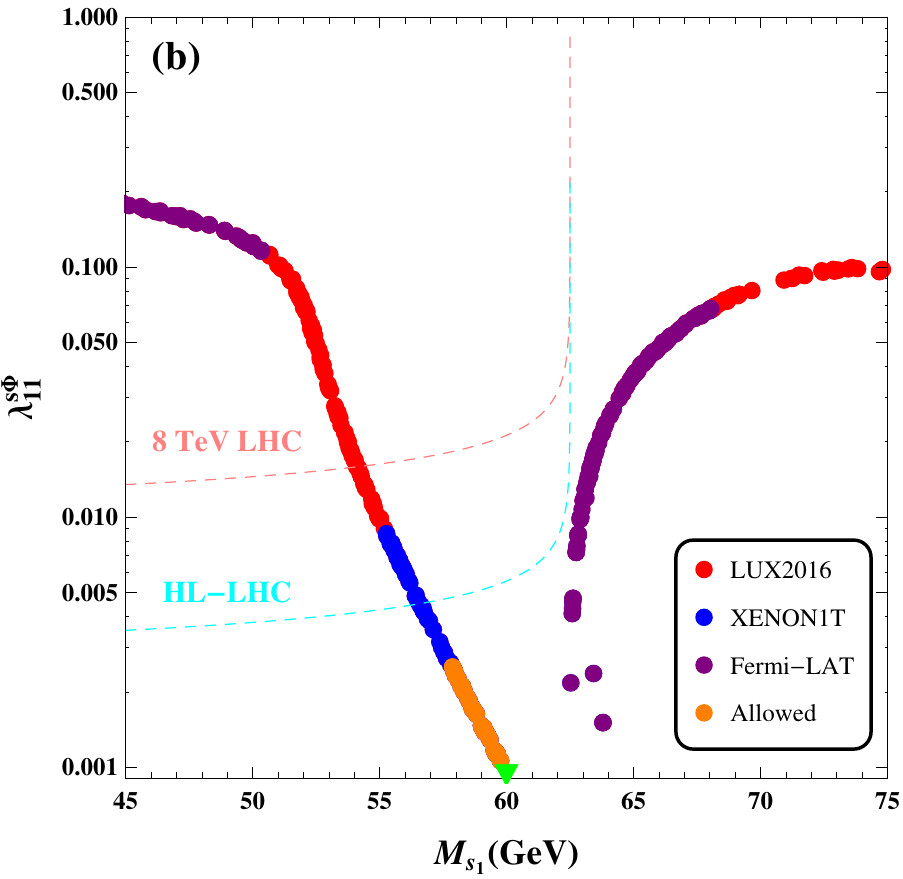}
\end{center}
\caption{Distributions of BR$_{\mbox{\tiny inv}}$ (a) and $\lambda^{s\Phi}_{11}$ (b) as a function of $M_{s_1}$ in the low mass region, using the same legends as in Fig.~\ref{DM_Para}. The dashed lines are current and expected upper bounds from LHC.
\label{DM_inv}}
\end{figure}

For $M_{s_1}>M_h/2$, it is challenging to probe $s_1$ DM with mono-jet signatures through the Higgs-portal at LHC \cite{Craig:2014lda}. For $M_{s_1}<M_h/2$, the new channel $h\to s_1^*s_1$ is kinematically opened, and contributes to invisible decays of the Higgs boson. The direct searches for invisible Higgs decays by LHC set an upper bound on the branching ratio $\mbox{BR}_{\mbox{\tiny inv}}$ of 0.28 in the weak boson fusion (WBF) channel~\cite{Aad:2015txa,Chatrchyan:2014tja} and 0.75 in the $Zh$ associated production channel~\cite{Chatrchyan:2014tja,Aad:2014iia}. Alternatively, a stronger bound comes from fitting to visible Higgs decays, i.e., $\mbox{BR}_{\mbox{\tiny inv}}<0.23$~\cite{Khachatryan:2014jba}, at $8~\TeV$ LHC. In principle, the WBF channel has the capability to probe the invisible branching ratio down to about $0.02$ at the high luminosity LHC (HL-LHC)~\cite{Bernaciak:2014pna}. The decay width of $h\to s_1^*s_1$ in the type II radiative seesaw reads:
\begin{equation}
\Gamma(h\to s_1^*s_1)=\frac{(\lambda^{s\Phi}_{11})^2 v^2}{16\pi M_h}\left(1-\frac{4M^2_{s_1}}{M^2_h}\right)^{1/2},
\end{equation}
so that the invisible branching ratio is calculated as $\mbox{BR}_{\mbox{\tiny inv}}=\Gamma_{\mbox{\tiny inv}}/(\Gamma_{\mbox{\tiny inv}}+\Gamma_{\mbox{\tiny SM}})$ with $\Gamma_{\mbox{\tiny SM}}=4.07~\MeV$ at $M_{h}=125~\GeV$~\cite{Djouadi:2005gi}. It is obvious that the invisible Higgs decay is strongly correlated with direct detection in the low mass region, since $\lambda^{s\Phi}_{11}$ and $M_{s_1}$ are the only two common variables in both processes~\cite{Baek:2014jga}.

The scatter plots of $\mbox{BR}_{\mbox{\tiny inv}}$ and $\lambda^{s\Phi}_{11}$ are presented in Fig.~\ref{DM_inv} as a function of $M_{s_1}$. For $M_{s_1}\lesssim52~\GeV$, $\mbox{BR}_{\mbox{\tiny inv}}$ is totally dominant, while for  $52~\GeV\lesssim M_{s_1}\lesssim62.5~\GeV$, $\mbox{BR}_{\mbox{\tiny inv}}$ decreases dramatically as $M_{s_1}$ increases. Currently, the $8~\TeV$ LHC has excluded $M_{s_1}\lesssim54~\GeV$, which is less stringent than the LUX2016 limit. The HL-LHC will be capable of excluding $M_{s_1}\lesssim57~\GeV$, which will be less stringent than the XENON1T limit. Therefore, we can always employ constraints from direct detections instead of invisible Higgs decays.

\section{LHC Signatures}\label{SEC:SG}

After our systematic study on dark matter properties in Sec. \ref{SEC:DM}, we now embark on the analysis of possible LHC signatures. As the benchmark points in Table \ref{Tab:BP} are on the safe side of current constraints from DM, we will employ them to illustrate multi-lepton signatures at LHC. To simulate signals and corresponding SM backgrounds, we generate the {\tt UFO} \cite{Degrande:2011ua} model file by {\tt FeynRules} \cite{Christensen:2008py}. The parton level events are produced with {\tt MadGraph5\_aMC@NLO} \cite{Alwall:2011uj} using the {\tt NNPDF2.3} \cite{Ball:2012cx} LO parton distribution function set, and pass through {\tt Pythia6} \cite{Sjostrand:2006za} to  include showering and hadronization. {\tt Delphes3}~\cite{Ovyn:2009tx} is then used for detector simulation and {\tt MadAnalysis5} \cite{Conte:2012fm} for analysis. The identification of $b$-jets is performed with a tagging efficiency of $70\%$, a mis-tagging rate of $10\%$ for $c$-jets and $1\%$ for light-flavor jets, respectively \cite{Chatrchyan:2012jua}.

\begin{figure}[!htbp]
\begin{center}
\includegraphics[width=0.45\linewidth]{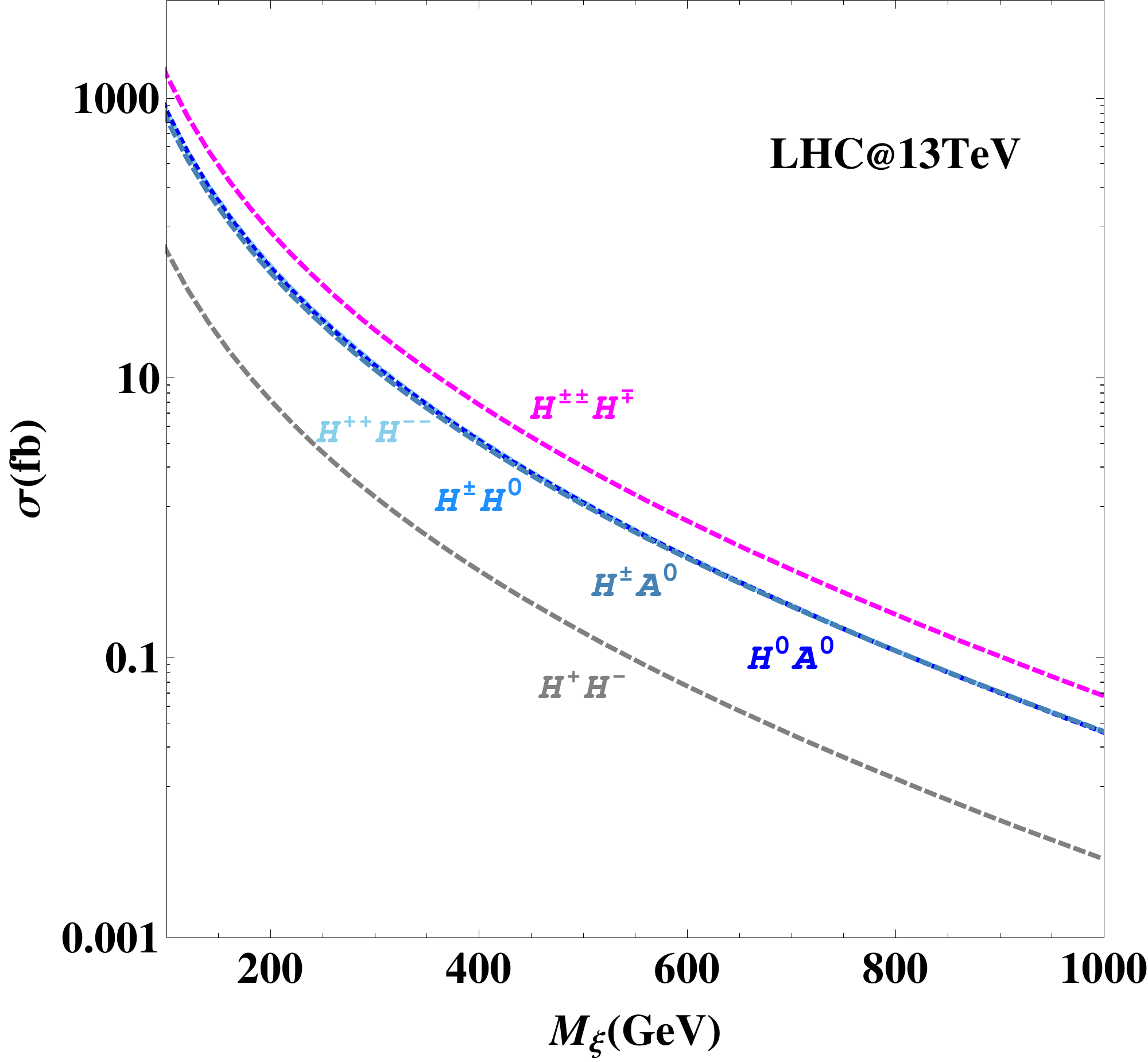}
\includegraphics[width=0.45\linewidth]{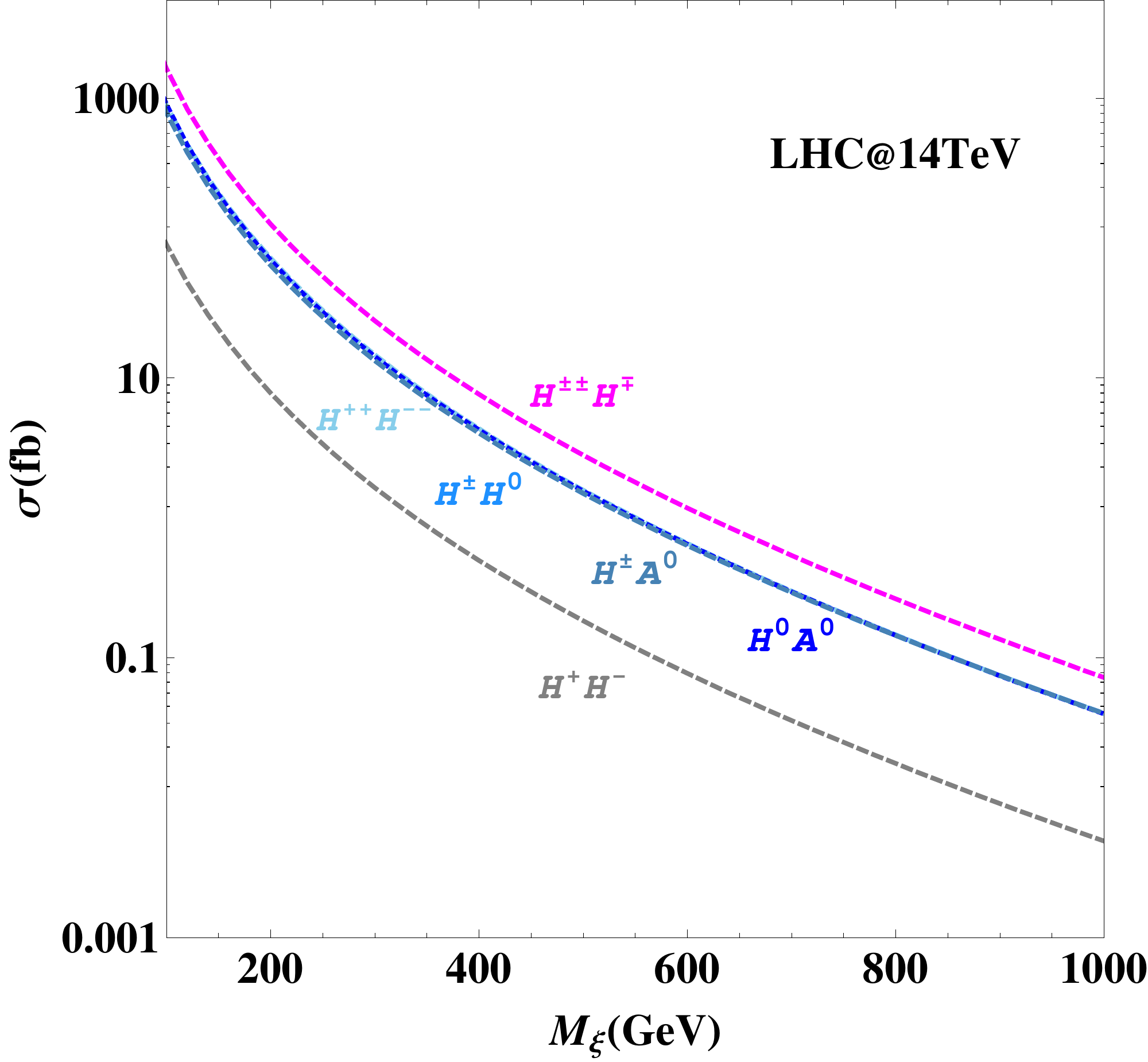}
\end{center}
\caption{Cross sections for pair and associated production of scalar triplet $\xi$ with a degenerate mass at LHC.}
\label{Production}
\end{figure}

In this work, we focus on new decay channels of the scalar triplet at LHC, e.g., $H^{++}\to E^+E^+$ with the subsequent decay $E^+\!\to \ell^+ s_1$. The production cross sections for pair and associated production of the scalar triplet are shown in Fig. \ref{Production}, which range from $1~\pb$ to $0.01~\fb$ in the mass interval $100-1000~\GeV$ at 13~TeV LHC, and become slightly bigger at 14~TeV. The production of $H^{++}H^{--}$ and $H^{\pm\pm}H^{\mp}$ will lead respectively to signatures of four-lepton and tri-lepton with a large missing transverse energy ($\cancel{E}_T$), due to the existence of the $s_1$ DM in the final states. With a larger cross section of $H^{\pm\pm}H^{\mp}$ than $H^{++}H^{--}$, the tri-lepton signature actually becomes a ``golden channel" in the canonical type II seesaw for the discovery of scalar triplet in its leptonic decay channels~\cite{Hpp:ph}. We expect the same to happen in the new decay channels.

Searches for four-lepton and tri-lepton plus $\cancel{E}_T$ signatures have recently been performed at 8~TeV LHC by CMS~\cite{Khachatryan:2014qwa,Chatrchyan:2014aea} and ATLAS~\cite{Aad:2014iza,Aad:2014nua}. These searches are usually based on simplified SUSY models, and thus their results must be taken with care when applying them to the type II radiative seesaw model which has different spectra, decay chains, and branching ratios. The analysis of Ref.~\cite{Fraser:2015mhb} for the four-lepton signature shows that the excluded region is only around $M_{\xi}\sim330~\GeV$ and $M_{\chi}\sim160~\GeV$, and the three benchmark points in Table \ref{Tab:BP} are out of this region. For the tri-lepton signature with less than 3 signal events after applying all cuts at 8~TeV LHC with $20~\fb^{-1}$ data, our three benchmark points are still consistent with current experimental limits at $95\%$ C.L.~\cite{Aad:2015eda}. It would be worthwhile to recast the SUSY search limits~\cite{Khachatryan:2014qwa,Chatrchyan:2014aea,Aad:2014iza,Aad:2014nua} on the type II radiative seesaw and examine their interplay with the DM constraints in the whole parameter space. We leave this for a future work.

\subsection{Decay Properties}

\begin{figure}[!htbp]
\begin{center}
\includegraphics[width=0.45\linewidth]{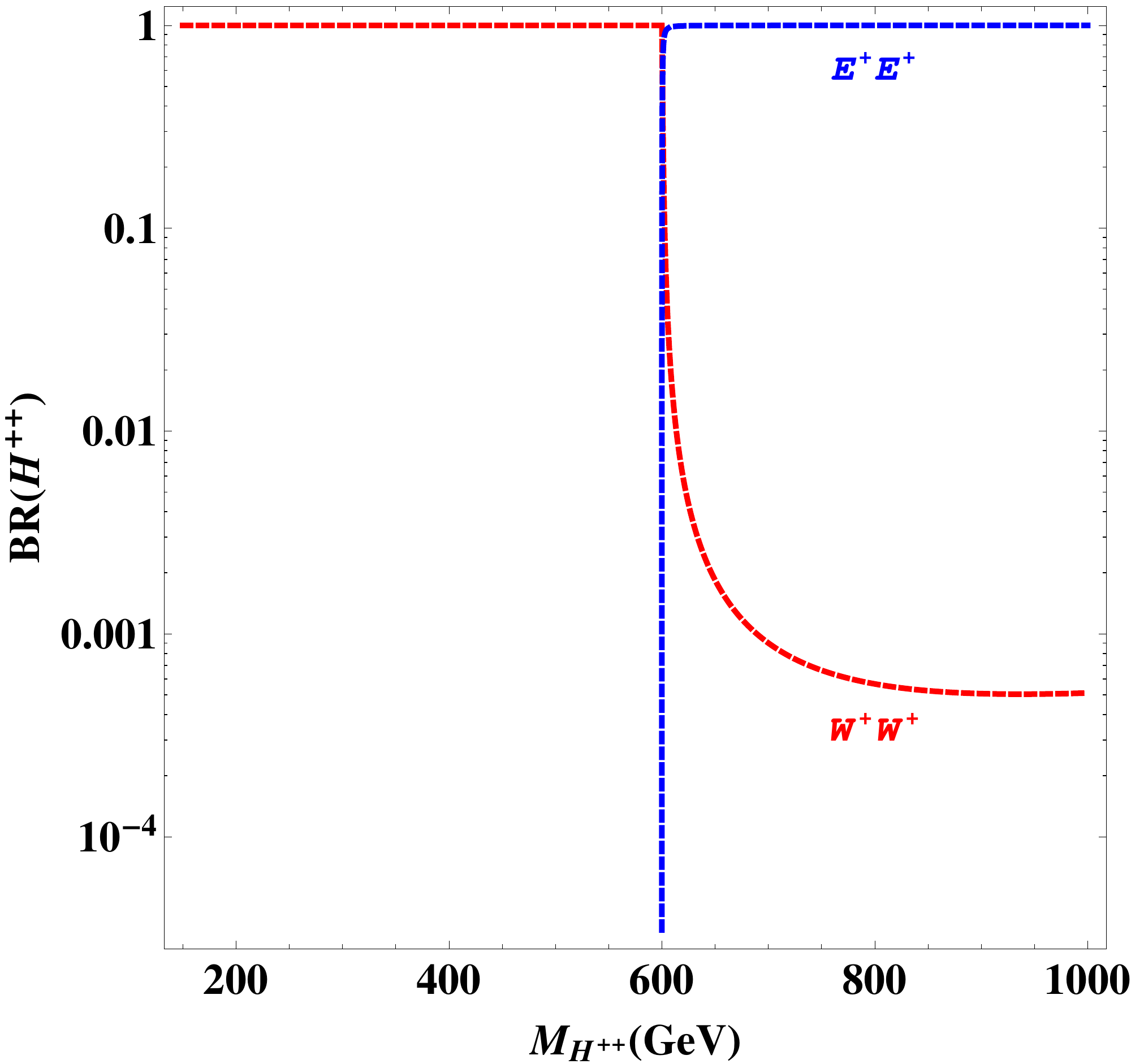}
\includegraphics[width=0.45\linewidth]{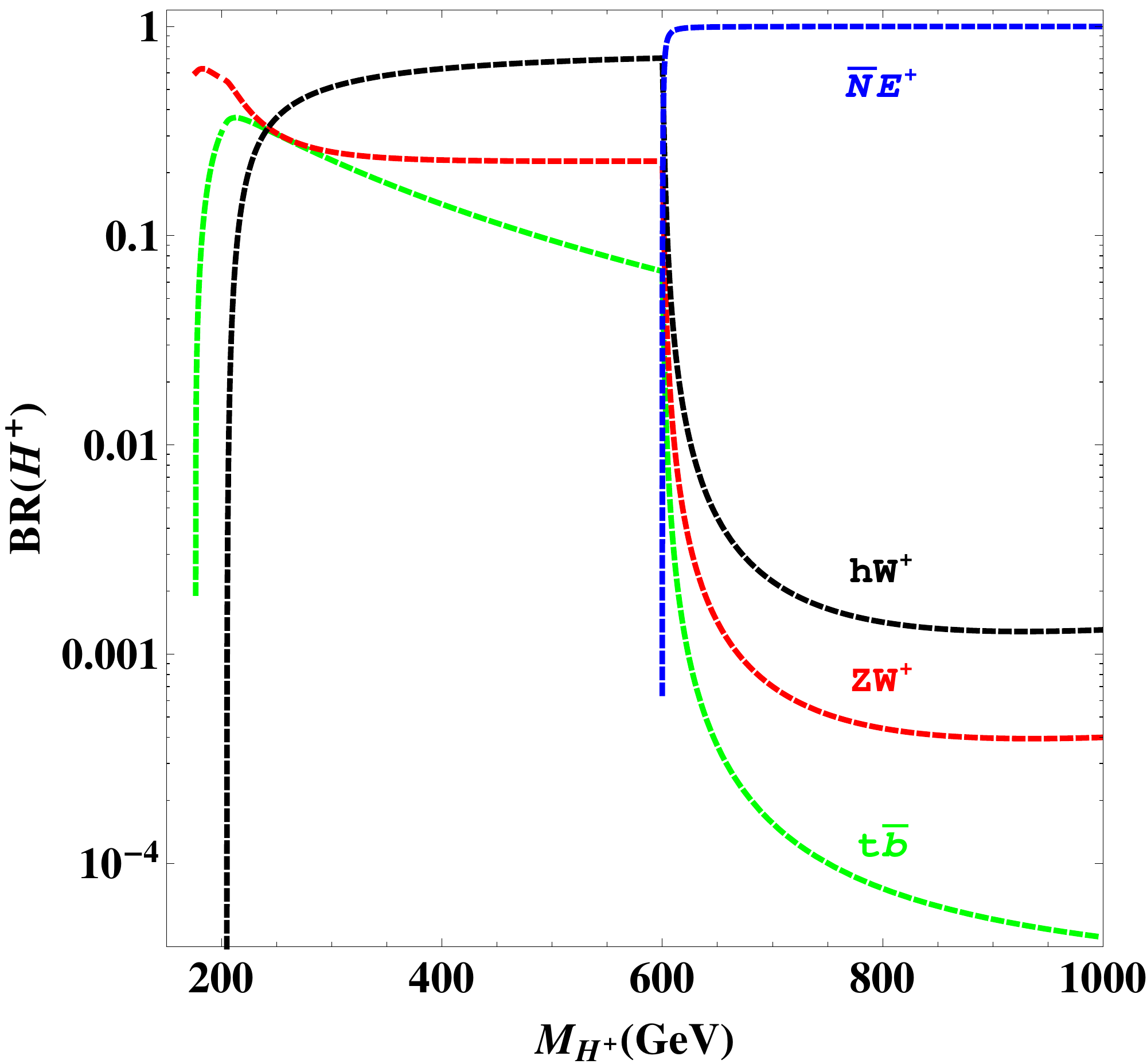}
\includegraphics[width=0.45\linewidth]{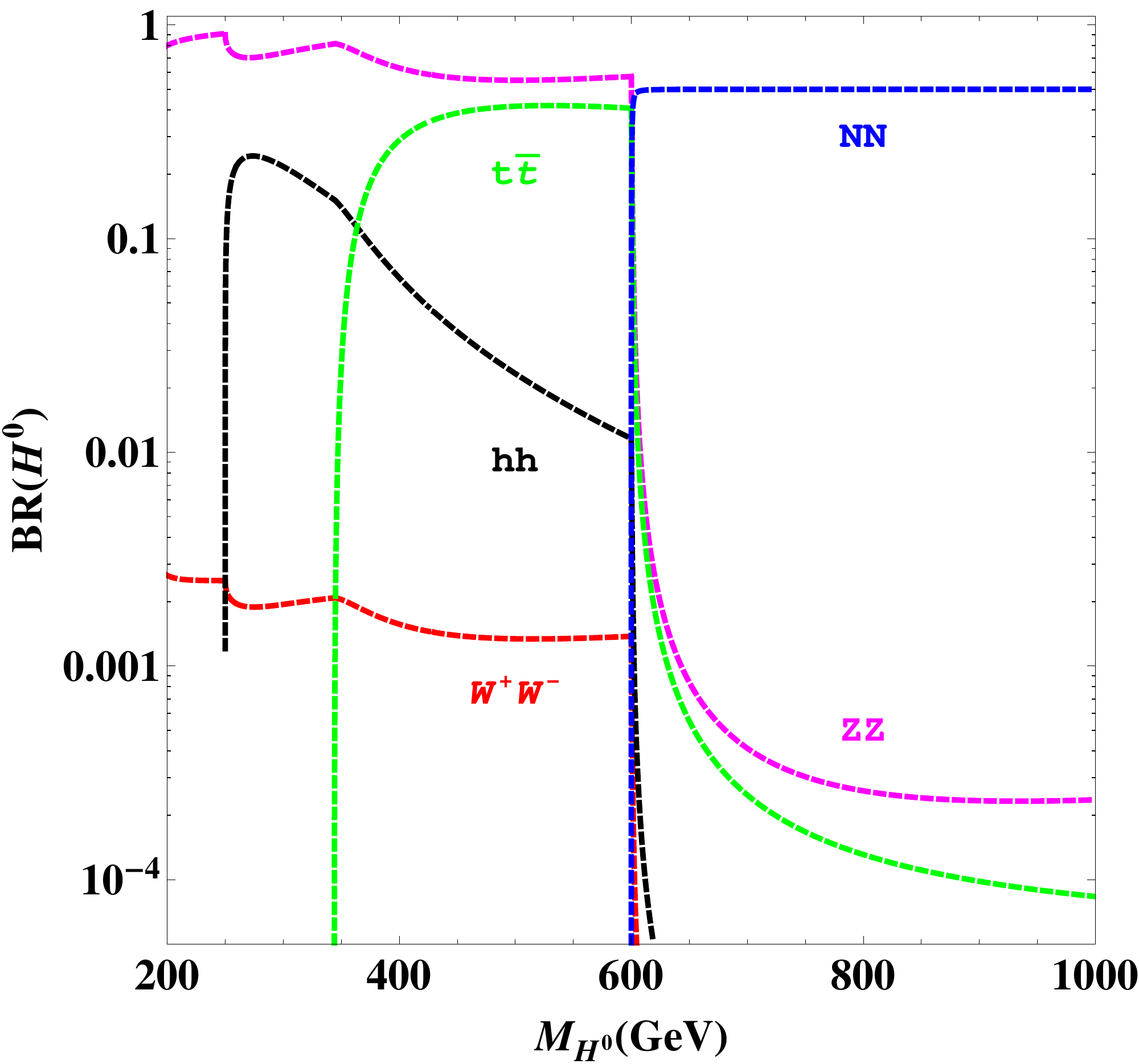}
\includegraphics[width=0.45\linewidth]{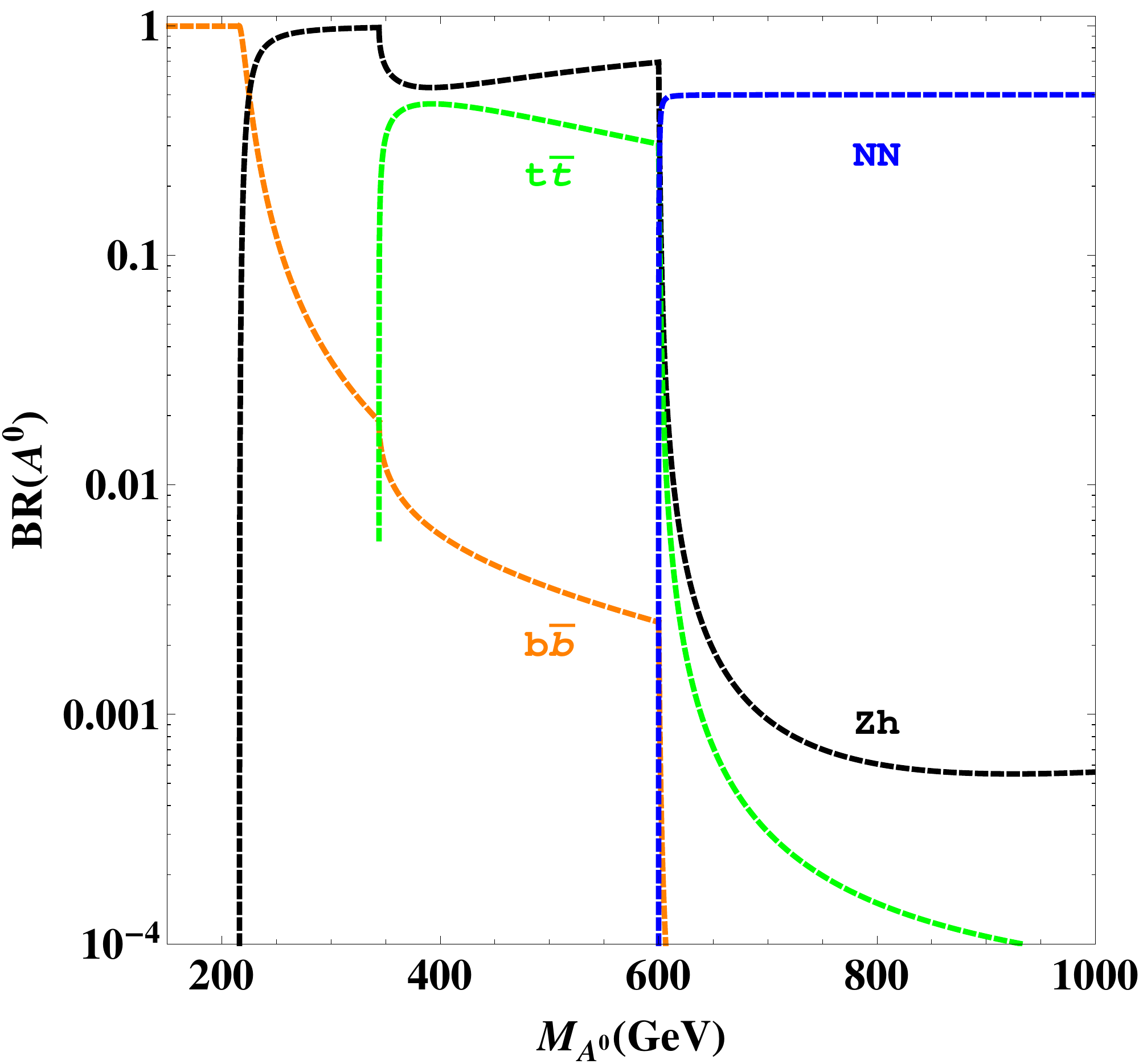}
\end{center}
\caption{Branching ratios of scalar triplet particles versus $M_{\xi}$ assuming $u=0.5~\GeV$, $M_{\chi}=300~\GeV$, and $z^{L,R}=1$.}
\label{Decay}
\end{figure}

Before detailed simulations on the signatures at LHC, we give a brief discussion on the decay properties of the scalar triplet $\xi$ and fermion doublet $\chi$. In Fig.~\ref{Decay}, we plot the branching ratios of the triplet particles as a function of $M_\xi$ by specifying $u=0.5~\GeV$, $M_{\chi}=300~\GeV$, and $z^{L,R}=1$, where the one-loop induced leptonic decays are not shown. The decays of the doubly-charged scalar $H^{++}$ are simple: when $M_{H^{++}}<2M_{\chi}$, the same-sign diboson channel $H^{++}\!\!\to W^+W^+$ dominates, and when $M_{H^{++}}>2M_{\chi}$, the decay $H^{++}\!\!\to E^+E^+$ takes over. For the singly-charged scalar $H^+$, one has $H^{+}\!\to hW^+,ZW^+,t\bar{b}$ when $M_{H^+}<2M_{\chi}$, and $H^{+}\!\to \bar{N}E^+$ when $M_{H^+}>2M_\chi$. For completeness, we also show the decay branching ratios of the neutral scalars $H^0$ and $A^0$, i.e., $H^0\to ZZ$ and $A^0\to Zh$ are dominant in the low mass region before the channels $H^0\to NN$ and $A^0\to NN$ are kinematically opened. In summary, when $M_{\xi}>2M_{\chi}$, the fermion decay channels, e.g., $H^{++}\to E^+E^+$ and $H^+\to \bar{N}E^+$, are dominant.

The fermion doublet $\chi$ can only decay into the SM leptons $F_L$ and inert scalars $s_a$ via the Yukawa coupling $x$. At our benchmark points shown in Table \ref{Tab:BP}, we have the mass order $M_{s_1}<M_{\chi}<M_{s_2}$, and thus the decay channels are simply $E^+\to \ell^+ s_1$ and $N\to \nu_\ell s_1^*$. Note that both decay products in $N\to \nu_\ell s_1^*$ are invisible at colliders, and there could be tight constraints from the mono-jet signature when $N$ is produced through the Drell-Yan process.

\subsection{Four-Lepton Signature}

The four-lepton signature is a good channel to probe doubly-charged scalars $H^{\pm\pm}$, mainly because of its clean SM background. It can only come from the $H^{++}H^{--}$ pair production with subsequent decays, $H^{\pm\pm}\to E^\pm E^\pm$ and $E^\pm\to \ell^\pm s_1^{(*)}$:
\begin{equation}
pp\to H^{++}H^{--} \to E^+E^+ E^-E^- \to \ell^+\ell^+\ell^-\ell^- + \cancel{E}_T,
\end{equation}
where $\ell=e,~\mu$ for collider simulations. To achieve a clean background, we concentrate on the final states without opposite-sign same-flavor (OSSF0) pair $\ell^+\ell^-$  as CMS~\cite{Khachatryan:2014qwa,Chatrchyan:2014aea} did for the four-lepton signature. The dominant sources of background are di-bosons ($WZ,~ZZ,~WW$), tri-bosons ($VVV$ with $V=W,~Z$), top pair ($t\bar{t}$), and top+boson (mainly from $t\bar{t}V$) with leptonic decays of $W,~Z$. The signals at the three benchmark points and their backgrounds are simulated at 13 (14)~TeV LHC with an integrated luminosity of $100~\fb^{-1}$. We adopt the same selection criteria as CMS \cite{Chatrchyan:2014aea} for a more realistic simulation.

\begin{figure}[!htbp]
\begin{center}
\includegraphics[width=0.33\linewidth]{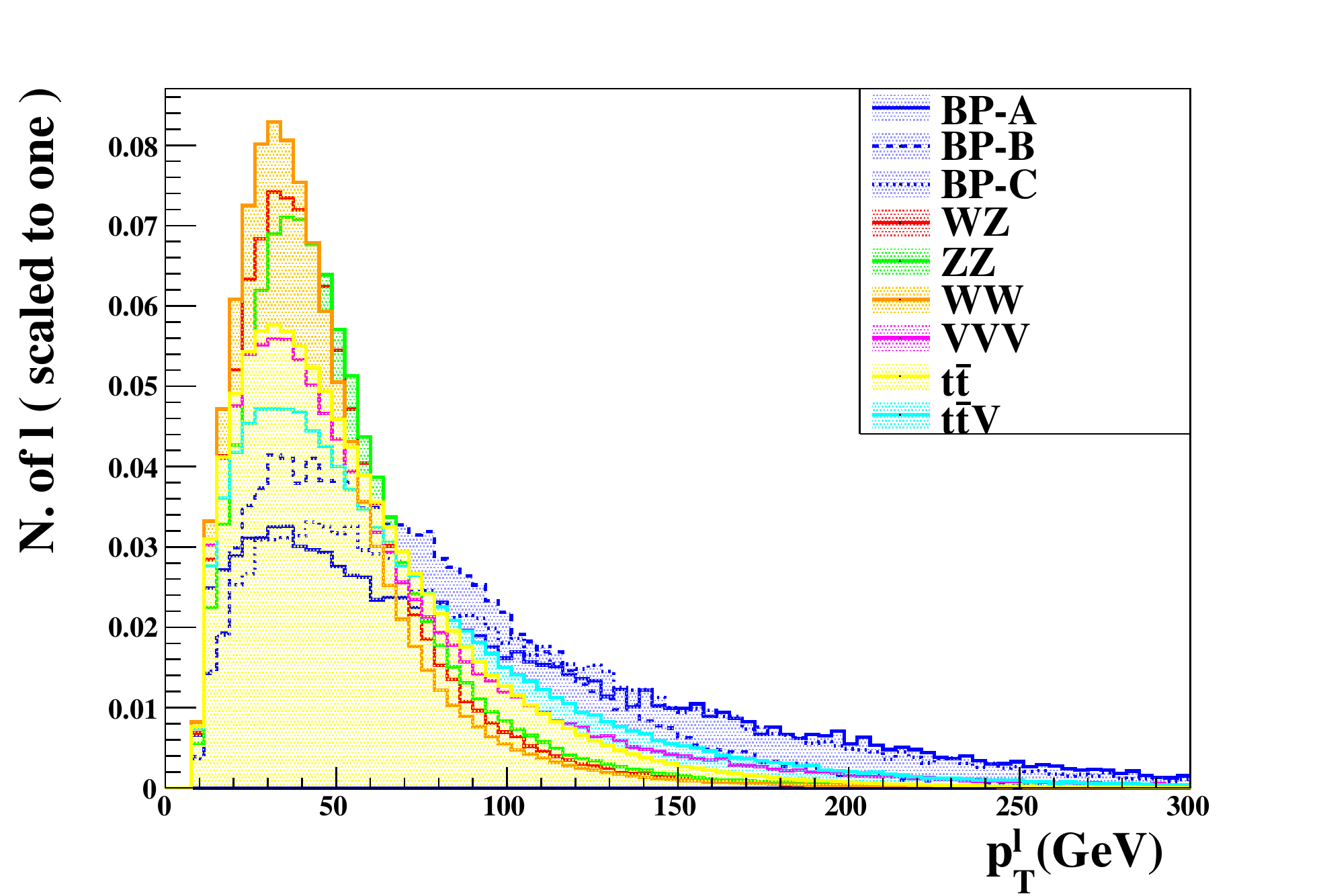}
\includegraphics[width=0.33\linewidth]{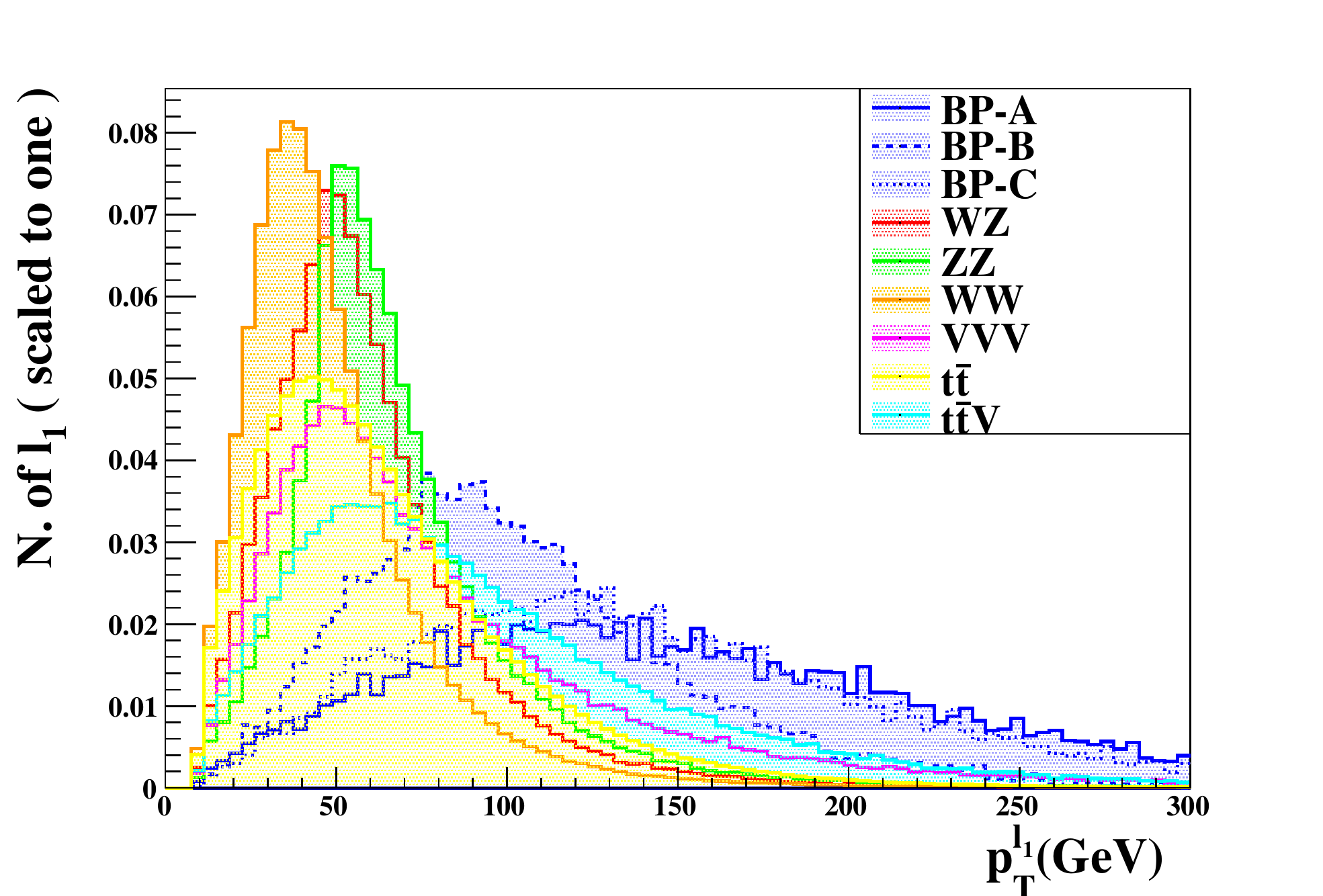}
\includegraphics[width=0.33\linewidth]{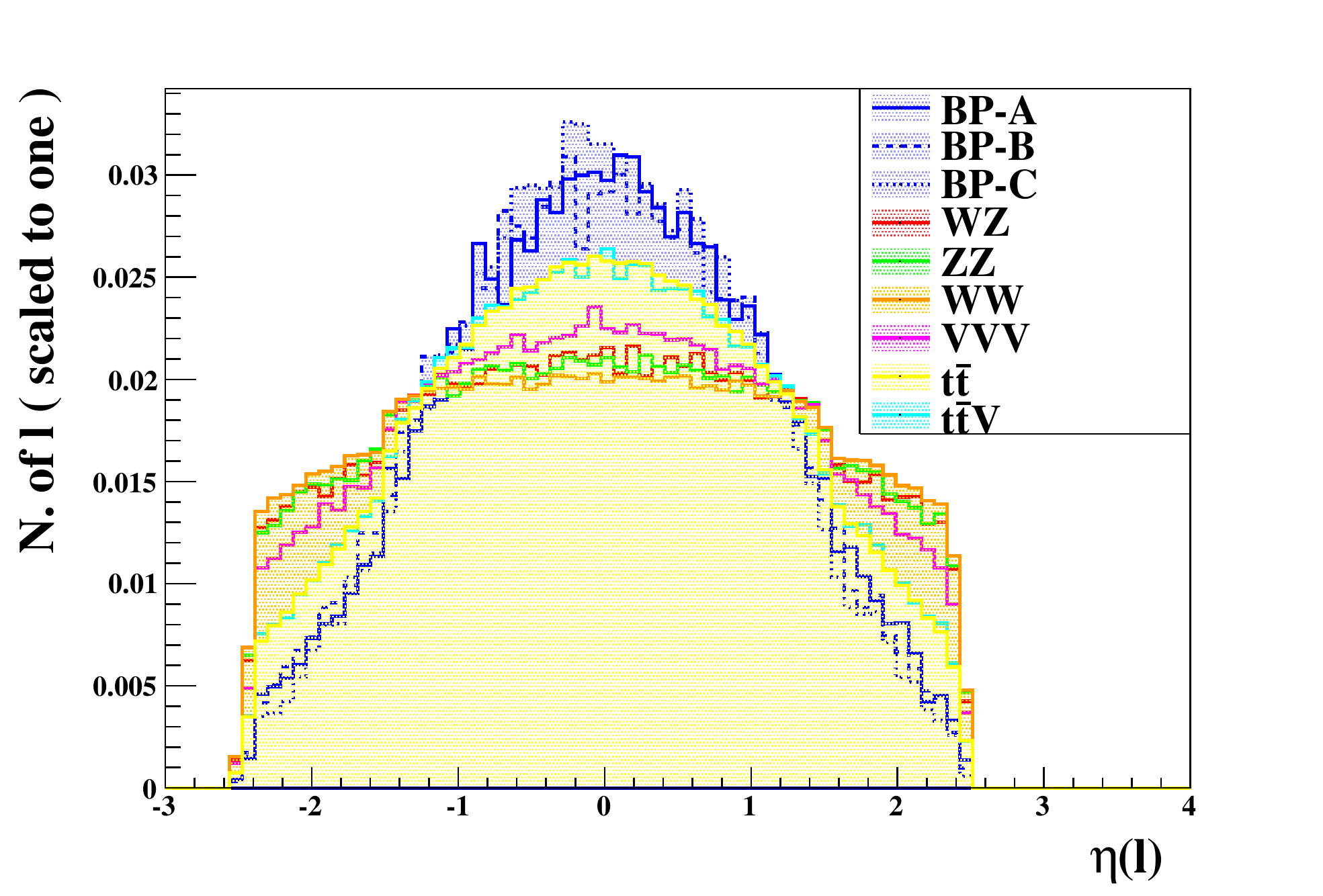}
\end{center}
\caption{Distributions of $p_T^\ell$, $p_T^{\ell_1}$, and $\eta(\ell)$ at $13~\TeV$ LHC for the four-lepton signature.}
\label{DIS_4l}
\end{figure}

We start with some basic cuts:
\begin{equation}\label{cut1}
p^\ell_T>10~\GeV, ~p^{\ell_1}_T>20~\GeV, ~|\eta(\ell)|<2.4,
\end{equation}
where $p^{\ell_1}_T$ denotes the transverse momentum of the most energetic one among four charged leptons. In Fig.~\ref{DIS_4l}, the distributions of $p^\ell_T$, $p^{\ell_1}_T$, and $\eta(\ell)$ at $13~\TeV$ are shown, and the results at $14~\TeV$ are similar. To reduce the background from semi-leptonic decays of heavy quarks, we also apply the lepton isolation criterion: $\sum_i p^i_T<0.15\,p_T^\ell$, where the sum is over all objects within a cone of radius $\Delta R=0.3$ around the lepton direction but excludes the lepton itself. Then we apply the following cuts to select the desired OSSF0 four-lepton events:
\begin{eqnarray}
N(\ell)=4,&\quad & N(b)=0,\\
N(e^+e^-)=0, &\quad & N(\mu^+\mu^-)=0.\label{cuts_NN}
\end{eqnarray}
Here, the cut on the number of $b$-jet mainly aims to reduce the $t\bar{t}$ and $t\bar{t}V$ backgrounds. In Table~\ref{TAB:4l}, we show the cut-flow for the four-lepton signature at the benchmark points and the dominant backgrounds. Our results are in agreement with Ref.~\cite{Fraser:2015mhb} and CMS~\cite{Chatrchyan:2014aea}. For the four-lepton events, the backgrounds are totally dominated by $ZZ$ after the basic cuts. The requirement of OSSF0 is then sufficient to suppress all backgrounds to a negligible level. We have about $17.0~(20.5)$, $4.79~(5.05)$, $2.11~(2.63)$ signal events at $13~(14)~\TeV$ LHC for the three benchmark points, respectively.

\begin{table}[!htbp]
\begin{tabular}{|c|c|c|c|c|c|}
\hline
Channels & No Cuts & Basic cuts in Eq.~(\ref{cut1})
& $N(\ell) = 4$ & $N(b) = 0$ & Cuts in Eq.~(\ref{cuts_NN})
\\
\hline
BP-A         & 173 (205)         & 170 (201)   & 54 (62) & 51 (59) & 6.3 (7.6)\\
BP-B         & 155 (184)         & 146 (174)   & 40 (44) & 38 (41) & 4.8 (5.1)\\
BP-C         & 62 (75)           & 60 (73)     & 18 (22) & 17 (21) & 2.1 (2.6)\\
\hline
$WZ$         & $3.60~(3.98)\cdot 10^4$     & $3.16~(3.45)\cdot 10^4$ & 0 (0) & 0 (0)  & 0 (0) \\
$ZZ$         & 4220 (4666)       & 3884~(4254)   & 782 (838) & 772 (826) & 0 (0)\\
$WW$         & $3.06~(3.36)\cdot 10^5$  & $2.26~(2.46)\cdot 10^5$   & 0 (0)  & 0 (0) & 0 (0)\\
$VVV$        & 145 (163)         & 133 (149)   & 5.61 (5.95)   & 5.50 (5.81)    & 0 (0)\\
$t\bar{t}$   & $2.27~(2.69)\cdot 10^6$ & $1.80~(2.11)\cdot 10^6$ & 0 (0)   & 0 (0)   & 0 (0)\\
$t\bar{t}V$  & 520 (604)         & 473 (549)   & 27.8 (32.9) & 5.06 (6.17)   & 0 (0)\\
\hline
\end{tabular}
\caption{Cut-flow for four-lepton signature at three benchmark points and dominant backgrounds at $13~(14)~\TeV$ LHC with an integrated luminosity of $100~\fb^{-1}$.}
\label{TAB:4l}
\end{table}

\subsection{Tri-Lepton Signature}

The tri-lepton signature is regarded as the golden channel for the scalar triplet particles, since the cross section for the $H^{\pm\pm}H^\mp$ associated production is about twice as large as the $H^{++}H^{--}$ pair production for degenerate masses~\cite{Hpp:ph}. The signature follows dominantly from $H^{\pm\pm}H^{\mp}$ production and subsequent decays, $H^{\pm\pm}\to E^\pm E^\pm$, $H^{\mp}\to E^\mp N$ and $E^\pm\to\ell^\pm s_1^{(*)}$, $N\to\nu_\ell s_1^{(*)}$:
\begin{equation}
pp\to H^{\pm\pm}H^{\mp} \to E^\pm E^\pm E^\mp N \to \ell^\pm\ell^\pm\ell^\mp+\cancel{E}_T.
\end{equation}
When simulating the four-lepton signature, we found that about half number of four-lepton events are actually detected as tri-lepton ones. Hence, in our following analysis for the tri-lepton signature, we consider contributions from both $H^{\pm\pm}H^{\mp}$ and $H^{++}H^{--}$ production. The two signatures also suffer similar SM backgrounds.

\begin{figure}[!htbp]
\begin{center}
\includegraphics[width=0.33\linewidth]{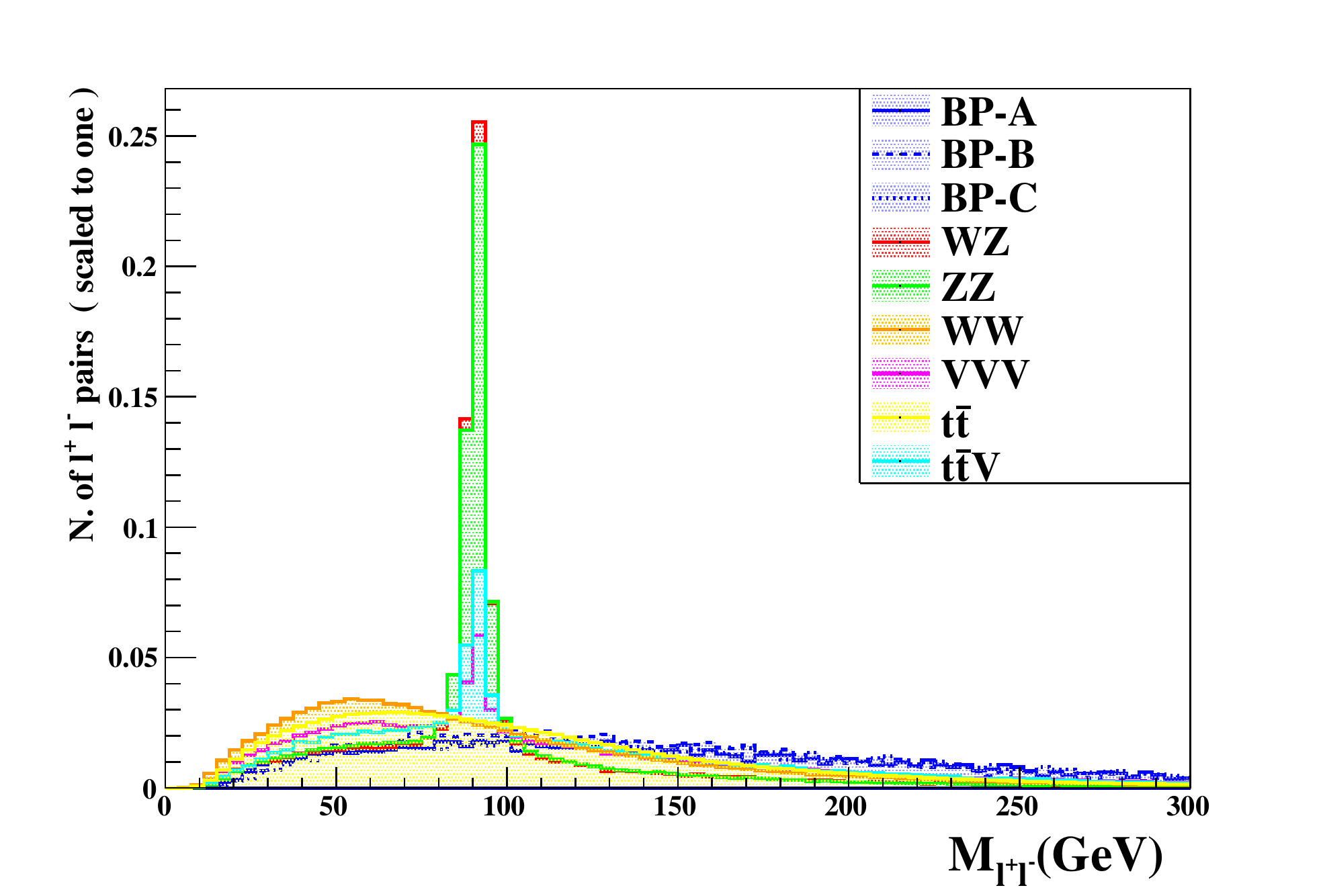}
\includegraphics[width=0.33\linewidth]{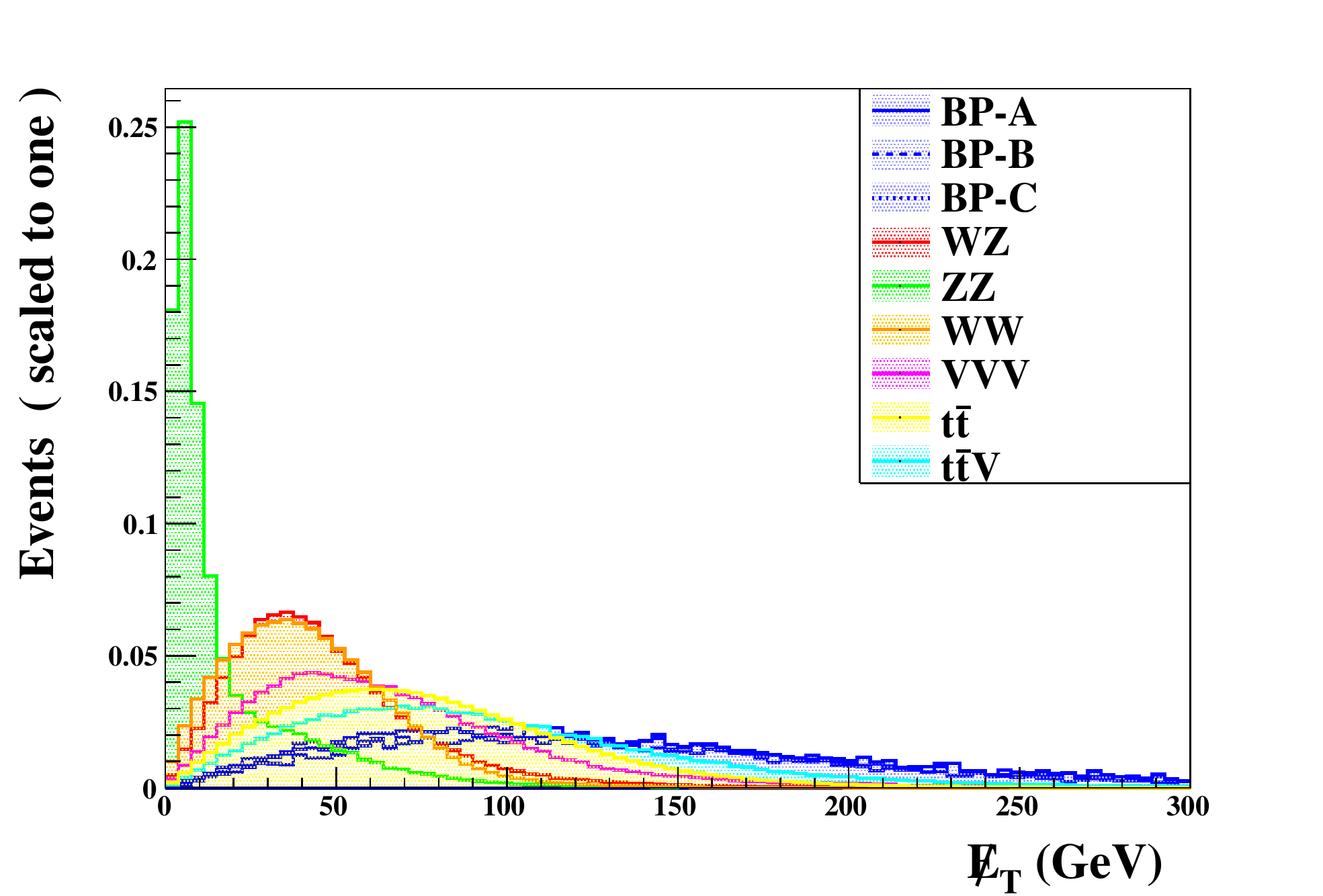}
\includegraphics[width=0.33\linewidth]{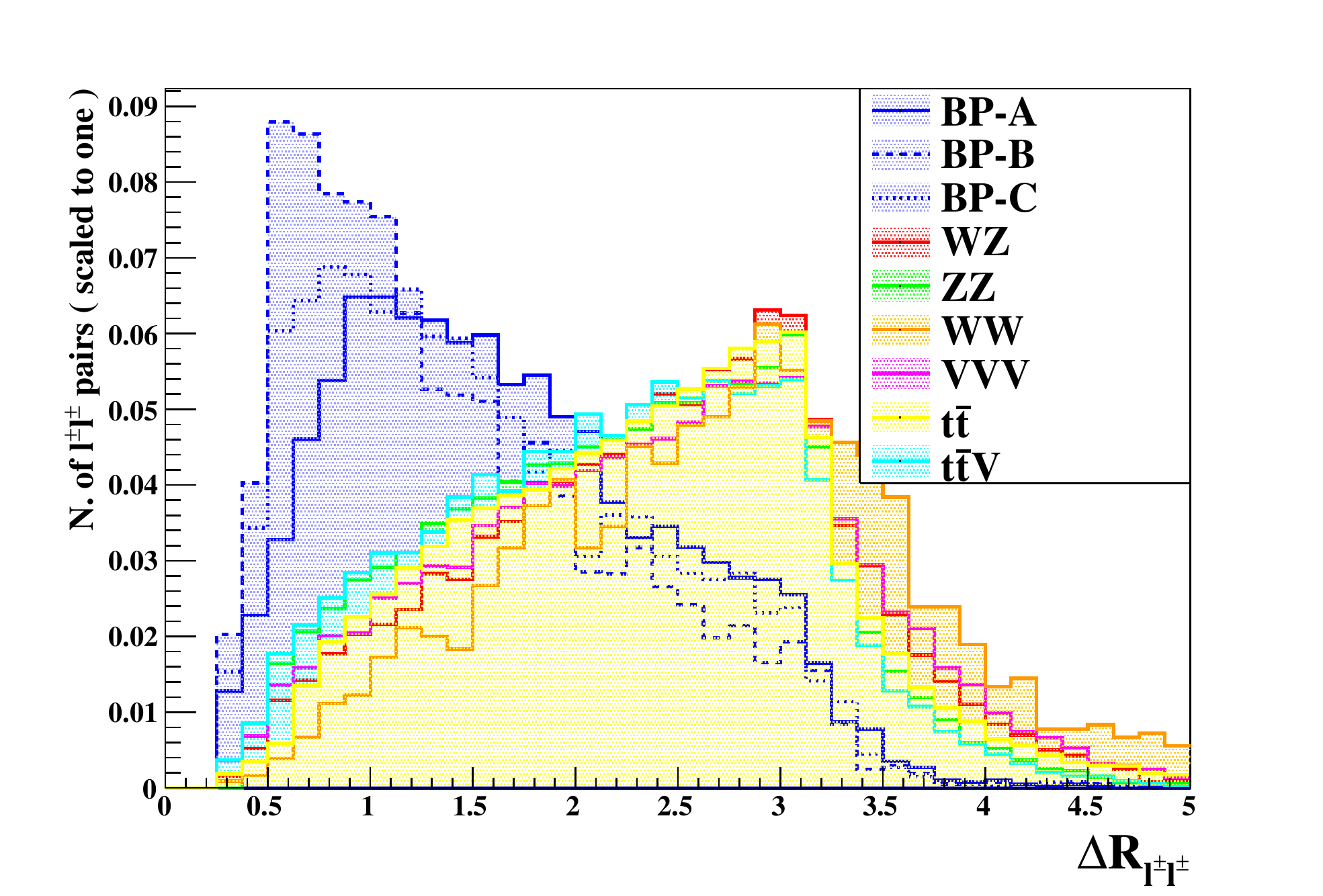}
\end{center}
\caption{Distributions of events in $M_{\ell^+\ell^-}$, $\cancel{E}_T$, and $\Delta R_{\ell^\pm\ell^\pm}$ at 13~TeV LHC for the tri-lepton signature.}
\label{DIS_3l}
\end{figure}

Again, we start with the basic cuts in Eq.~(\ref{cut1}). Then we select the tri-lepton events by adopting the cuts:
\begin{equation}
\label{cut_tri}
N(\ell)=3,\quad N(b)=0.
\end{equation}
At this stage, the dominant backgrounds are from $WZ$ and $ZZ$. In principle, we can apply the same cuts as CMS~\cite{Chatrchyan:2014aea} or ATLAS~\cite{Aad:2014iza,Aad:2014nua} to further reduce the backgrounds. But even if we choose the OSSF0 signal region, there are still a lot of backgrounds survived. This is mainly because that the experimental cuts~\cite{Khachatryan:2014qwa,Chatrchyan:2014aea,Aad:2014iza,Aad:2014nua} are particularly designed for hunting SUSY particles instead of scalar triplet particles in this model. To get some hints about further cuts, we show in Fig.~\ref{DIS_3l} the distributions of events in $M_{\ell^+\ell^-}$, $\cancel{E}_T$, and $\Delta R_{\ell^\pm\ell^\pm}$ at 13~TeV LHC. (The results at 14~TeV are similar.) It is clear that the dominant backgrounds $WZ$ and $ZZ$ have a sharp peak around $M_Z$ in the distribution of $M_{\ell^+\ell^-}$ while the signals do not. We therefore make a $Z$-veto cut to delete events with $85~\GeV<M_{\ell^+\ell^-}<95~\GeV$. In the tri-lepton signature at our benchmark points, both neutrino $\nu_\ell$ and DM $s_1$ lead to a large missing transverse energy $\cancel{E}_T$, which suggests the cut, $\cancel{E}_T>150~\GeV$. Furthermore, the same-sign lepton pair ($\ell^\pm\ell^\pm$) from $H^{\pm\pm}$ decays tends to be closer to each other than in the backgrounds, a cut on the separation between the two same-sign leptons, $\Delta R_{\ell^\pm\ell^\pm}<2$, is appropriate according to Fig.~\ref{DIS_3l}.

Table~\ref{TAB:3l} shows the cut-flow for the tri-lepton signature at the benchmark points together with backgrounds. The cuts we employed here are efficient enough in preserving the signal while suppressing the backgrounds. At the three benchmark points, we have about 92.54~(116.83), 28.64~(34.65), and 18.99~(22.86) events at 13~(14) TeV with only about 2 background events.

\begin{table}[!htbp]
\begin{tabular}{|c|c|c|c|c|c|c|}
\hline
Channels & No Cuts & Basic cuts in Eq.~(\ref{cut1}) & Cuts in Eq.~(\ref{cut_tri})
& $Z$-veto & $\cancel{E}_T\! > 150~\GeV $ &  $ \Delta R_{\ell^\pm\ell^\pm} <2$ \\
\hline
BP-A           & 562 (665)         & 543 (640)   & 215 (253) & 144 (169) &57.0 (65.2)  & 48.1 (55.2) \\
BP-B           & 501 (597)         & 472 (561)   & 172 (202) & 114 (133) &31.1 (38.0)   & 28.6 (34.7) \\
BP-C           & 202 (245)         & 194 (235)   & 76.9 (91.3)   & 52.4 (63.8)   &21.7 (25.9) & 19.0 (22.9) \\
\hline
$WZ$           & $3.60~(3.98)\cdot 10^4$ & $3.16~(3.45)\cdot 10^4$ & 8492 (9012) & 836 (932) &16.2 (20.7) & 1.08 (0.79) \\
$ZZ$           & 4220 (4666)       & 3884 (4254)   & 1218 (1311) & 119 (129) & 0 (0.23)& 0 (0.05) \\
$WW$           & $3.06~(3.36)\cdot 10^5$ & $2.26~(2.46)\cdot 10^5$ & 0.31 (0.67)& 0.31 (0.67) & 0 (0) & 0 (0) \\
$VVV$          & 145 (163)         & 133 (149)   & 40.5 (44.7)    & 19.7 (21.6)   & 1.17 (1.20)     & 0.35 (0.32) \\
$t\bar{t}$     & $2.27~(2.69)\cdot 10^6$ & $1.80~(2.11)\cdot 10^6$ & 36.4 (25.4) & 14.1 (9.76) & 0.91 (0)     & 0.45 (0) \\
$t\bar{t}V$    & 520 (604)         & 473 (549)   & 25.7 (30.0)   & 11.1 (12.9)   & 1.29 (1.40) & 0.51 (0.62) \\
\hline
\end{tabular}
\caption{Cut-flow for tri-lepton signature at three benchmark points and dominant backgrounds at 13~(14)~TeV LHC with an integrated luminosity of $100~\fb^{-1}$.}
\label{TAB:3l}
\end{table}

Before ending up this section, we summarize our simulation results on the four- and tri-lepton signatures at LHC. In Table \ref{Tab:SIM}, we list the survival numbers of signal events $S$ and background events $B$, as well as the statistical significance $S/\sqrt{S+B}$ after applying all cuts. The background for the four-lepton signature is very clean, but in the meanwhile only about $2-5$ signal events survive, leading to a significance less than $3\sigma$. The tri-lepton signal events are about $6-9$ times larger, and the corresponding significance could reach about $5\sigma$, albeit there are a few background events. Therefore, we could conclude that the tri-lepton signature is more promising than the four-lepton one.

\begin{table}[!htbp]
\begin{tabular}{|c|c|c|c|c|c|c|}
\hline
\raisebox{-0.5ex}[0pt]{~Benchmark~}& \multicolumn{3}{c|}{Four-Lepton}& \multicolumn{3}{c|}{Tri-Lepton}\\
\cline{2-7}
\raisebox{0.5ex}[0pt]{points}& $S$ & $B$ & $S/\sqrt{S+B}$ & $S$ & $B$ & $S/\sqrt{S+B}$\\
\hline
BP-A  &~6.30 (7.60)~&            &~2.50 (2.76)~&~48.2 (55.2)~&            &~6.77 (7.31)~ \\
\cline{1-2}\cline{4-5}\cline{7-7}
BP-B  & 4.79 (5.05) &~0 (0)~& 2.19 (2.25) & 28.6 (34.7) &~2.39 (1.78)~& 5.14 (5.74)\\
\cline{1-2}\cline{4-5}\cline{7-7}
BP-C  & 2.11 (2.63) &            & 1.45 (1.62) & 19.0 (22.9) &            & 4.11 (4.61)\\
\hline
\end{tabular}
\caption{Testability of four- and tri-lepton signatures at 13~(14)~TeV LHC. The four-lepton signature contains only the OSSF0 final states.}
\label{Tab:SIM}
\end{table}

\section{Conclusion}\label{SEC:CL}

We have made a detailed analysis on the testability of the type II radiative seesaw that relates neutrino mass and dark matter at one-loop level. After incorporating the constraints from lepton flavor violation and collider searches, we focused on the dark matter properties and LHC signatures. We found that introduction of a heavier singlet scalar $s_2$ can greatly enlarge the allowed parameter space compared to the minimal case with one $s_1$ DM particle. And the upcoming experiments of direct detection, XENON1T, and indirect detection, CTA, have the capability of probing a large portion of the enlarged parameter space. By considering the combined constraints from relic density, direct and indirect detection, and invisible Higgs decays, we found three possible regions of $M_{s_1}$ that can satisfy all these constraints at present and even in the future: (1) the Higgs resonance region $M_{s_1}\sim M_h/2$, (2) the Higgs region $M_{s_1}\sim M_h$, and (3) the coannihilation region $M_{s_2}\sim M_{s_1}$.

Based on the above results on dark matter properties, we have chosen three benchmark points to illustrate possible collider signatures of the model. We have concentrated on new decay channels of the charged scalars, i.e., $H^{++}\to E^+E^+$ and $H^+\to \bar{N}E^+$, with subsequent decays $E^+\to \ell^+ s_1$ and $N \to \nu_\ell s_1^*$. Our simulations show that the four- and tri-lepton signatures arising from $H^{++}H^{--}$ and $H^{\pm\pm}H^{\mp}$ production respectively are quite promising to be probed at LHC, and in particular the tri-lepton signature can reach $\sim 5\sigma$ significance at 13 or 14~TeV LHC with $100~\fb^{-1}$ data.

\section*{Acknowledgement}
This work was supported in part by the Grants No. NSFC-11025525, No. NSFC-11575089 and by the CAS Center for Excellence in Particle Physics (CCEPP). The numerical analysis was done with the HPC Cluster of SKLTP/ITP-CAS.



\begin{thebibliography}{000}

\bibitem{Weinberg:1979sa}
  S.~Weinberg,
  Phys.\ Rev.\ Lett.\  {\bf 43}, 1566 (1979).

\bibitem{Ma:1998dn}
  E.~Ma,
  Phys.\ Rev.\ Lett.\  {\bf 81}, 1171 (1998)
  [hep-ph/9805219].

\bibitem{type1}
  P.~Minkowski,
  Phys.\ Lett.\ B {\bf 67}, 421 (1977),
T.~Yanagida, in  {\em Proceedings of the Workshop on Unified
Theories and Baryon Number in the Universe}, eds. O. Sawada et al.,
(KEK Report 79-18, Tsukuba, 1979), p. 95;
M.~Gell-Mann, P.~Ramond, R.~Slansky, in {\em Supergravity},
eds. P. Van Niewenhuizen et al., (North-Holland, 1979), p. 315;
S.~Glashow, in {\em Quarks and Leptons}, Carg\`{e}ses,
eds. M. L\'{e}vy et al., (Plenum, 1980), p. 707;
  R.~N.~Mohapatra and G.~Senjanovic,
  Phys.\ Rev.\ Lett.\  {\bf 44}, 912 (1980).

\bibitem{type2}
  M.~Magg and C.~Wetterich,
  Phys.\ Lett.\ B {\bf 94}, 61 (1980).
  T.~P.~Cheng and L.~F.~Li,
  Phys.\ Rev.\ D {\bf 22}, 2860 (1980).
  J.~Schechter and J.~W.~F.~Valle,
  Phys.\ Rev.\ D {\bf 22}, 2227 (1980).
  G.~Lazarides, Q.~Shafi and C.~Wetterich,
  Nucl.\ Phys.\ B {\bf 181}, 287 (1981).
  R.~N.~Mohapatra and G.~Senjanovic,
  Phys.\ Rev.\ D {\bf 23}, 165 (1981).

\bibitem{type3}
  R.~Foot, H.~Lew, X.~G.~He and G.~C.~Joshi,
  Z.\ Phys.\ C {\bf 44}, 441 (1989).

\bibitem{Jungman:1995df}
  G.~Jungman, M.~Kamionkowski and K.~Griest,
  Phys.\ Rept.\  {\bf 267}, 195 (1996)
  [hep-ph/9506380].
  G.~Bertone,
  Nature {\bf 468}, 389 (2010)
  [arXiv:1011.3532 [astro-ph.CO]].

\bibitem{radiative}
  A.~Zee,
  Phys.\ Lett.\ B {\bf 93}, 389 (1980)
  Erratum: [Phys.\ Lett.\ B {\bf 95}, 461 (1980)].
  L.~Wolfenstein,
  Nucl.\ Phys.\ B {\bf 175}, 93 (1980).
  A.~Zee,
  Nucl.\ Phys.\ B {\bf 264}, 99 (1986).
  K.~S.~Babu,
  Phys.\ Lett.\ B {\bf 203}, 132 (1988).
  K.~S.~Babu and E.~Ma,
  Phys.\ Rev.\ Lett.\  {\bf 61}, 674 (1988).

\bibitem{Boucenna:2014zba}
  S.~M.~Boucenna, S.~Morisi and J.~W.~F.~Valle,
  Adv.\ High Energy Phys.\  {\bf 2014}, 831598 (2014)
  [arXiv:1404.3751 [hep-ph]].

\bibitem{Ma:2006km}
  E.~Ma,
  Phys.\ Rev.\ D {\bf 73}, 077301 (2006)
  [hep-ph/0601225].

\bibitem{1loop}
  J.~Kubo, E.~Ma and D.~Suematsu,
  Phys.\ Lett.\ B {\bf 642}, 18 (2006)
  [hep-ph/0604114].
  P.~H.~Gu and U.~Sarkar,
  Phys.\ Rev.\ D {\bf 77}, 105031 (2008)
  [arXiv:0712.2933 [hep-ph]].
  D.~Aristizabal Sierra, J.~Kubo, D.~Restrepo, D.~Suematsu and O.~Zapata,
  Phys.\ Rev.\ D {\bf 79}, 013011 (2009)
  [arXiv:0808.3340 [hep-ph]].
  E.~Ma and D.~Suematsu,
  Mod.\ Phys.\ Lett.\ A {\bf 24}, 583 (2009)
  [arXiv:0809.0942 [hep-ph]].
  S.~Kanemura, T.~Nabeshima and H.~Sugiyama,
  Phys.\ Lett.\ B {\bf 703}, 66 (2011)
  [arXiv:1106.2480 [hep-ph]].
  S.~Kanemura, O.~Seto and T.~Shimomura,
  Phys.\ Rev.\ D {\bf 84}, 016004 (2011)
  [arXiv:1101.5713 [hep-ph]].
  S.~Kanemura and H.~Sugiyama,
  Phys.\ Rev.\ D {\bf 86}, 073006 (2012)
  [arXiv:1202.5231 [hep-ph]].
  F.~Bonnet, M.~Hirsch, T.~Ota and W.~Winter,
  JHEP {\bf 1207}, 153 (2012)
  [arXiv:1204.5862 [hep-ph]].
  H.~Okada and T.~Toma,
  Phys.\ Rev.\ D {\bf 86}, 033011 (2012)
  [arXiv:1207.0864 [hep-ph]].
  P.~S.~B.~Dev and A.~Pilaftsis,
  Phys.\ Rev.\ D {\bf 86}, 113001 (2012)
  [arXiv:1209.4051 [hep-ph]].
  D.~Schmidt, T.~Schwetz and T.~Toma,
  Phys.\ Rev.\ D {\bf 85}, 073009 (2012)
  [arXiv:1201.0906 [hep-ph]].
  M.~Hirsch, R.~A.~Lineros, S.~Morisi, J.~Palacio, N.~Rojas and J.~W.~F.~Valle,
  JHEP {\bf 1310}, 149 (2013)
  [arXiv:1307.8134 [hep-ph]].
  S.~S.~C.~Law and K.~L.~McDonald,
  JHEP {\bf 1309}, 092 (2013)
  [arXiv:1305.6467 [hep-ph]].
  S.~Kanemura, T.~Matsui and H.~Sugiyama,
  Phys.\ Lett.\ B {\bf 727}, 151 (2013)
  [arXiv:1305.4521 [hep-ph]].
  D.~Restrepo, O.~Zapata and C.~E.~Yaguna,
  JHEP {\bf 1311}, 011 (2013)
  [arXiv:1308.3655 [hep-ph]].
  V.~Brdar, I.~Picek and B.~Radovcic,
  Phys.\ Lett.\ B {\bf 728}, 198 (2014)
  [arXiv:1310.3183 [hep-ph]].
  H.~Okada and K.~Yagyu,
  Phys.\ Rev.\ D {\bf 90}, 035019 (2014)
  [arXiv:1405.2368 [hep-ph]].
  W.~Wang and Z.~L.~Han,
  Phys.\ Rev.\ D {\bf 92}, 095001 (2015)
  [arXiv:1508.00706 [hep-ph]].
  R.~Longas, D.~Portillo, D.~Restrepo and O.~Zapata,
  JHEP {\bf 1603}, 162 (2016)
  [arXiv:1511.01873 [hep-ph]].
  R.~Adhikari, D.~Borah and E.~Ma,
  Phys.\ Lett.\ B {\bf 755}, 414 (2016)
  [arXiv:1512.05491 [hep-ph]].
  H.~Okada and Y.~Orikasa,
  arXiv:1512.06687 [hep-ph].
  A.~Ibarra, C.~E.~Yaguna and O.~Zapata,
  Phys.\ Rev.\ D {\bf 93}, 035012 (2016)
  [arXiv:1601.01163 [hep-ph]].
  R.~Ding, Z.~L.~Han, Y.~Liao and X.~D.~Ma,
  Eur.\ Phys.\ J.\ C {\bf 76}, 204 (2016)
  [arXiv:1601.02714 [hep-ph]].
  A.~Ahriche, K.~L.~McDonald, S.~Nasri and I.~Picek,
  Phys.\ Lett.\ B {\bf 757}, 399 (2016)
  [arXiv:1603.01247 [hep-ph]].
  C.~Kownacki and E.~Ma,
  Phys.\ Lett.\ B {\bf 760}, 59 (2016)
  [arXiv:1604.01148 [hep-ph]].
  E.~Ma, N.~Pollard, O.~Popov and M.~Zakeri,
  arXiv:1605.00991 [hep-ph].
  T.~Nomura, H.~Okada and Y.~Orikasa,
  arXiv:1605.02601 [hep-ph].
  T.~Nomura and H.~Okada,
  arXiv:1606.09055 [hep-ph].
  P.~H.~Gu, E.~Ma and U.~Sarkar,
  arXiv:1608.02118 [hep-ph].

\bibitem{2loop}
  E.~Ma,
  Phys.\ Lett.\ B {\bf 662}, 49 (2008)
  [arXiv:0708.3371 [hep-ph]].
  S.~Kanemura, T.~Nabeshima and H.~Sugiyama,
  Phys.\ Rev.\ D {\bf 85}, 033004 (2012)
  [arXiv:1111.0599 [hep-ph]].
  S.~Baek, P.~Ko, H.~Okada and E.~Senaha,
  JHEP {\bf 1409}, 153 (2014)
  [arXiv:1209.1685 [hep-ph]].
  Y.~Kajiyama, H.~Okada and K.~Yagyu,
  Nucl.\ Phys.\ B {\bf 874}, 198 (2013)
  [arXiv:1303.3463 [hep-ph]].
  Y.~Kajiyama, H.~Okada and T.~Toma,
  Phys.\ Rev.\ D {\bf 88}, 015029 (2013)
  [arXiv:1303.7356].
  S.~Baek, H.~Okada and T.~Toma,
  JCAP {\bf 1406}, 027 (2014)
  [arXiv:1312.3761 [hep-ph]].
  H.~Okada,
  arXiv:1404.0280 [hep-ph].
  S.~Kanemura, T.~Matsui and H.~Sugiyama,
  Phys.\ Rev.\ D {\bf 90}, 013001 (2014)
  [arXiv:1405.1935 [hep-ph]].
  M.~Aoki and T.~Toma,
  JCAP {\bf 1409}, 016 (2014)
  [arXiv:1405.5870 [hep-ph]].
  H.~Okada, T.~Toma and K.~Yagyu,
  Phys.\ Rev.\ D {\bf 90}, 095005 (2014)
  [arXiv:1408.0961 [hep-ph]].
  D.~Aristizabal Sierra, A.~Degee, L.~Dorame and M.~Hirsch,
  JHEP {\bf 1503}, 040 (2015)
  [arXiv:1411.7038 [hep-ph]].
  H.~Okada,
  arXiv:1503.04557 [hep-ph].
  S.~Kashiwase, H.~Okada, Y.~Orikasa and T.~Toma,
  Int.\ J.\ Mod.\ Phys.\ A {\bf 31}, 1650121 (2016)
  [arXiv:1505.04665 [hep-ph]].
  H.~Okada and Y.~Orikasa,
  Phys.\ Rev.\ D {\bf 93}, 013008 (2016)
  [arXiv:1509.04068 [hep-ph]].
  R.~Ding, Z.~L.~Han, Y.~Liao and W.~P.~Xie,
  JHEP {\bf 1605}, 030 (2016)
  [arXiv:1601.06355 [hep-ph]].
  T.~Nomura and H.~Okada,
  Phys.\ Lett.\ B {\bf 756}, 295 (2016)
  [arXiv:1601.07339 [hep-ph]].
  T.~Nomura, H.~Okada and Y.~Orikasa,
  arXiv:1602.08302 [hep-ph].
  C.~Bonilla, E.~Ma, E.~Peinado and J.~W.~F.~Valle,
  arXiv:1607.03931 [hep-ph].
  T.~Nomura and H.~Okada,
  arXiv:1607.04952 [hep-ph].

\bibitem{3loop}
  L.~M.~Krauss, S.~Nasri and M.~Trodden,
  Phys.\ Rev.\ D {\bf 67}, 085002 (2003)
  [hep-ph/0210389].
  K.~Cheung and O.~Seto,
  Phys.\ Rev.\ D {\bf 69}, 113009 (2004)
  [hep-ph/0403003].
  M.~Aoki, S.~Kanemura and O.~Seto,
  Phys.\ Rev.\ Lett.\  {\bf 102}, 051805 (2009)
  [arXiv:0807.0361 [hep-ph]].
  M.~Gustafsson, J.~M.~No and M.~A.~Rivera,
  Phys.\ Rev.\ Lett.\  {\bf 110}, 211802 (2013)
  Erratum: [Phys.\ Rev.\ Lett.\  {\bf 112}, 259902 (2014)]
  [arXiv:1212.4806 [hep-ph]].
  J.~N.~Ng and A.~de la Puente,
  Phys.\ Lett.\ B {\bf 727}, 204 (2013)
  [arXiv:1307.2606 [hep-ph]].
  Y.~Kajiyama, H.~Okada and K.~Yagyu,
  JHEP {\bf 1310}, 196 (2013)
  [arXiv:1307.0480 [hep-ph]].
  A.~Ahriche, C.~S.~Chen, K.~L.~McDonald and S.~Nasri,
  Phys.\ Rev.\ D {\bf 90}, 015024 (2014)
  [arXiv:1404.2696 [hep-ph]].
  H.~Hatanaka, K.~Nishiwaki, H.~Okada and Y.~Orikasa,
  Nucl.\ Phys.\ B {\bf 894}, 268 (2015)
  [arXiv:1412.8664 [hep-ph]].
  K.~Nishiwaki, H.~Okada and Y.~Orikasa,
  Phys.\ Rev.\ D {\bf 92}, 093013 (2015)
  [arXiv:1507.02412 [hep-ph]].
  H.~Okada and K.~Yagyu,
  Phys.\ Rev.\ D {\bf 93}, 013004 (2016)
  [arXiv:1508.01046 [hep-ph]].
  A.~Ahriche, K.~L.~McDonald and S.~Nasri,
  JHEP {\bf 1602}, 038 (2016)
  [arXiv:1508.02607 [hep-ph]].
  S.~Kanemura, K.~Nishiwaki, H.~Okada, Y.~Orikasa, S.~C.~Park and R.~Watanabe,
  arXiv:1512.09048 [hep-ph].
  H.~Okada and K.~Yagyu,
  Phys.\ Lett.\ B {\bf 756}, 337 (2016)
  [arXiv:1601.05038 [hep-ph]].
  P.~Ko, T.~Nomura, H.~Okada and Y.~Orikasa,
  Phys.\ Rev.\ D {\bf 94}, 013009 (2016)
  [arXiv:1602.07214 [hep-ph]].
  T.~Nomura, H.~Okada and Y.~Orikasa,
  Phys.\ Rev.\ D {\bf 93}, 113008 (2016)
  [arXiv:1603.04631 [hep-ph]].
  D.~Cherigui, C.~Guella, A.~Ahriche and S.~Nasri,
  arXiv:1605.03640 [hep-ph].
  T.~Nomura, H.~Okada and N.~Okada,
  arXiv:1608.02694 [hep-ph].

\bibitem{4loop}
  T.~Nomura and H.~Okada,
  Phys.\ Lett.\ B {\bf 755}, 306 (2016)
  [arXiv:1601.00386 [hep-ph]].
  T.~Nomura and H.~Okada,
  arXiv:1601.04516 [hep-ph].

\bibitem{Krauss:1988zc}
  L.~M.~Krauss and F.~Wilczek,
  Phys.\ Rev.\ Lett.\  {\bf 62}, 1221 (1989);
  B.~Batell,
  Phys.\ Rev.\ D {\bf 83}, 035006 (2011)
  [arXiv:1007.0045 [hep-ph]].

\bibitem{Ma:2015xla}
  E.~Ma,
  Phys.\ Rev.\ Lett.\  {\bf 115}, 011801 (2015)
  [arXiv:1502.02200 [hep-ph]].

\bibitem{Ma:2013mga}
  E.~Ma,
  Phys.\ Rev.\ Lett.\  {\bf 112}, 091801 (2014)
  [arXiv:1311.3213 [hep-ph]].

\bibitem{Fraser:2015mhb}
  S.~Fraser, C.~Kownacki, E.~Ma and O.~Popov,
  Phys.\ Rev.\ D {\bf 93}, 013021 (2016)
  [arXiv:1511.06375 [hep-ph]].

\bibitem{Bhattacharya:2015qpa}
  S.~Bhattacharya, N.~Sahoo and N.~Sahu,
  arXiv:1510.02760 [hep-ph].

  \bibitem{SVL}
  Y.~Chikashige, R.~N.~Mohapatra and R.~D.~Peccei,
  Phys.\ Lett.\ B {\bf 98}, 265 (1981).
  G.~B.~Gelmini and M.~Roncadelli,
  Phys.\ Lett.\ B {\bf 99}, 411 (1981).
  C.~S.~Aulakh and R.~N.~Mohapatra,
  Phys.\ Lett.\ B {\bf 119}, 136 (1982).
  J.~Schechter and J.~W.~F.~Valle,
  Phys.\ Rev.\ D {\bf 25}, 774 (1982).
  W.~Wang and Z.~L.~Han,
  arXiv:1605.00239 [hep-ph].

\bibitem{Agashe:2014kda}
  K.~A.~Olive {\it et al.} [Particle Data Group Collaboration],
  Chin.\ Phys.\ C {\bf 38}, 090001 (2014).

\bibitem{Kanemura:2012rs}
  S.~Kanemura and K.~Yagyu,
  Phys.\ Rev.\ D {\bf 85}, 115009 (2012)
  [arXiv:1201.6287 [hep-ph]].
\bibitem{Chun:2012jw}
  E.~J.~Chun, H.~M.~Lee and P.~Sharma,
  JHEP {\bf 1211}, 106 (2012)
  [arXiv:1209.1303 [hep-ph]].
\bibitem{Aoki:2012jj}
  M.~Aoki, S.~Kanemura, M.~Kikuchi and K.~Yagyu,
  Phys.\ Rev.\ D {\bf 87}, 015012 (2013)
  [arXiv:1211.6029 [hep-ph]].

\bibitem{Arhrib:2011uy}
  A.~Arhrib, R.~Benbrik, M.~Chabab, G.~Moultaka, M.~C.~Peyranere, L.~Rahili and J.~Ramadan,
  Phys.\ Rev.\ D {\bf 84}, 095005 (2011)
  [arXiv:1105.1925 [hep-ph]].
\bibitem{Bonilla:2015eha}
  C.~Bonilla, R.~M.~Fonseca and J.~W.~F.~Valle,
  Phys.\ Rev.\ D {\bf 92}, 075028 (2015)
  [arXiv:1508.02323 [hep-ph]].
\bibitem{Chabab:2015nel}
  M.~Chabab, M.~Capdequi-Peyran¨¨re and L.~Rahili,
  arXiv:1512.07280 [hep-ph].

\bibitem{Haba:2016zbu}
  N.~Haba, H.~Ishida, N.~Okada and Y.~Yamaguchi,
  arXiv:1601.05217 [hep-ph].

\bibitem{Das:2016bir}
  D.~Das and A.~Santamaria,
  arXiv:1604.08099 [hep-ph].

\bibitem{Akeroyd:2011zza}
  A.~G.~Akeroyd and H.~Sugiyama,
  Phys.\ Rev.\ D {\bf 84}, 035010 (2011)
  [arXiv:1105.2209 [hep-ph]].

\bibitem{Melfo:2011nx}
  A.~Melfo, M.~Nemevsek, F.~Nesti, G.~Senjanovic and Y.~Zhang,
  Phys.\ Rev.\ D {\bf 85}, 055018 (2012)
  [arXiv:1108.4416 [hep-ph]].

\bibitem{Aoki:2011pz}
  M.~Aoki, S.~Kanemura and K.~Yagyu,
  Phys.\ Rev.\ D {\bf 85}, 055007 (2012)
  [arXiv:1110.4625 [hep-ph]].

\bibitem{Akeroyd:2012nd}
  A.~G.~Akeroyd, S.~Moretti and H.~Sugiyama,
  Phys.\ Rev.\ D {\bf 85}, 055026 (2012)
  [arXiv:1201.5047 [hep-ph]].

\bibitem{Chun:2012zu}
  E.~J.~Chun and P.~Sharma,
  JHEP {\bf 1208}, 162 (2012)
  [arXiv:1206.6278 [hep-ph]].
  E.~J.~Chun and P.~Sharma,
  Phys.\ Lett.\ B {\bf 722}, 86 (2013)
  [arXiv:1301.1437 [hep-ph]].
  E.~J.~Chun and P.~Sharma,
  Phys.\ Lett.\ B {\bf 728}, 256 (2014)
  [arXiv:1309.6888 [hep-ph]].

\bibitem{Han:2015hba}
  Z.~L.~Han, R.~Ding and Y.~Liao,
  Phys.\ Rev.\ D {\bf 91}, 093006 (2015)
  [arXiv:1502.05242 [hep-ph]].
  Z.~L.~Han, R.~Ding and Y.~Liao,
  Phys.\ Rev.\ D {\bf 92}, 033014 (2015)
  [arXiv:1506.08996 [hep-ph]].

\bibitem{Aad:2012tfa}
  G.~Aad {\it et al.} [ATLAS Collaboration],
  Phys.\ Lett.\ B {\bf 716}, 1 (2012)
  [arXiv:1207.7214 [hep-ex]].

\bibitem{Chatrchyan:2012xdj}
  S.~Chatrchyan {\it et al.} [CMS Collaboration],
  Phys.\ Lett.\ B {\bf 716}, 30 (2012)
  [arXiv:1207.7235 [hep-ex]].

\bibitem{Aad:2015zhl}
  G.~Aad {\it et al.} [ATLAS and CMS Collaborations],
  Phys.\ Rev.\ Lett.\  {\bf 114}, 191803 (2015)
  [arXiv:1503.07589 [hep-ex]].

\bibitem{Chun:2003ej}
  E.~J.~Chun, K.~Y.~Lee and S.~C.~Park,
  Phys.\ Lett.\ B {\bf 566}, 142 (2003)
  [hep-ph/0304069].
  A.~G.~Akeroyd, M.~Aoki and H.~Sugiyama,
  Phys.\ Rev.\ D {\bf 79}, 113010 (2009)
  [arXiv:0904.3640 [hep-ph]].
  T.~Fukuyama, H.~Sugiyama and K.~Tsumura,
  JHEP {\bf 1003}, 044 (2010)
  [arXiv:0909.4943 [hep-ph]].


\bibitem{Ding:2014nga}
  R.~Ding, Z.~L.~Han, Y.~Liao, H.~J.~Liu and J.~Y.~Liu,
  Phys.\ Rev.\ D {\bf 89}, 115024 (2014)
  [arXiv:1403.2040 [hep-ph]].

\bibitem{MEG}
  J.~Adam {\it et al.} [MEG Collaboration],
  Phys.\ Rev.\ Lett.\  {\bf 110}, 201801 (2013)
  [arXiv:1303.0754 [hep-ex]].
  [The MEG Collaboration],
  arXiv:1605.05081 [hep-ex].

\bibitem{Hpp:ph}
  A.~G.~Akeroyd and M.~Aoki,
  Phys.\ Rev.\ D {\bf 72}, 035011 (2005)
  [hep-ph/0506176].
  T.~Han, B.~Mukhopadhyaya, Z.~Si and K.~Wang,
  Phys.\ Rev.\ D {\bf 76}, 075013 (2007)
  [arXiv:0706.0441 [hep-ph]].
  P.~Fileviez Perez, T.~Han, G.~y.~Huang, T.~Li and K.~Wang,
  Phys.\ Rev.\ D {\bf 78}, 015018 (2008)
  [arXiv:0805.3536 [hep-ph]].
  F.~del Aguila and J.~A.~Aguilar-Saavedra,
  Nucl.\ Phys.\ B {\bf 813}, 22 (2009)
  [arXiv:0808.2468 [hep-ph]].

\bibitem{ATLAS:2012hi}
  G.~Aad {\it et al.} [ATLAS Collaboration],
  Eur.\ Phys.\ J.\ C {\bf 72}, 2244 (2012)
  [arXiv:1210.5070 [hep-ex]].
  G.~Aad {\it et al.} [ATLAS Collaboration],
  JHEP {\bf 1503}, 041 (2015)
  [arXiv:1412.0237 [hep-ex]].

\bibitem{Chatrchyan:2012ya}
  S.~Chatrchyan {\it et al.} [CMS Collaboration],
  Eur.\ Phys.\ J.\ C {\bf 72}, 2189 (2012)
  [arXiv:1207.2666 [hep-ex]].
  CMS Collaboration,
  CMS-PAS-HIG-14-039.

\bibitem{Kanemura:2013vxa}
  S.~Kanemura, K.~Yagyu and H.~Yokoya,
  Phys.\ Lett.\ B {\bf 726}, 316 (2013)
  [arXiv:1305.2383 [hep-ph]].
  S.~Kanemura, M.~Kikuchi, K.~Yagyu and H.~Yokoya,
  Phys.\ Rev.\ D {\bf 90}, 115018 (2014)
  [arXiv:1407.6547 [hep-ph]].
  S.~Kanemura, M.~Kikuchi, H.~Yokoya and K.~Yagyu,
  PTEP {\bf 2015}, 051B02 (2015)
  [arXiv:1412.7603 [hep-ph]].

\bibitem{Aad:2014vma}
  G.~Aad {\it et al.} [ATLAS Collaboration],
  JHEP {\bf 1405}, 071 (2014)
  [arXiv:1403.5294 [hep-ex]].

\bibitem{Khachatryan:2014qwa}
  V.~Khachatryan {\it et al.} [CMS Collaboration],
  Eur.\ Phys.\ J.\ C {\bf 74}, 3036 (2014)
  [arXiv:1405.7570 [hep-ex]].

\bibitem{Christensen:2008py}
  N.~D.~Christensen and C.~Duhr,
  Comput.\ Phys.\ Commun.\  {\bf 180}, 1614 (2009)
  [arXiv:0806.4194 [hep-ph]].
  N.~D.~Christensen, P.~de Aquino, C.~Degrande, C.~Duhr, B.~Fuks, M.~Herquet, F.~Maltoni and S.~Schumann,
  Eur.\ Phys.\ J.\ C {\bf 71}, 1541 (2011)
  [arXiv:0906.2474 [hep-ph]].
  A.~Alloul, N.~D.~Christensen, C.~Degrande, C.~Duhr and B.~Fuks,
  Comput.\ Phys.\ Commun.\  {\bf 185}, 2250 (2014)
  [arXiv:1310.1921 [hep-ph]].

\bibitem{Belyaev:2012qa}
  A.~Belyaev, N.~D.~Christensen and A.~Pukhov,
  Comput.\ Phys.\ Commun.\  {\bf 184}, 1729 (2013)
  [arXiv:1207.6082 [hep-ph]].
  A.~Pukhov,
  hep-ph/0412191.

\bibitem{Belanger:2014vza}
  G.~B¨¦langer, F.~Boudjema, A.~Pukhov and A.~Semenov,
  Comput.\ Phys.\ Commun.\  {\bf 192}, 322 (2015)
  [arXiv:1407.6129 [hep-ph]].
  G.~Belanger, F.~Boudjema, A.~Pukhov and A.~Semenov,
  Comput.\ Phys.\ Commun.\  {\bf 185}, 960 (2014)
  [arXiv:1305.0237 [hep-ph]].
  G.~Belanger, F.~Boudjema, A.~Pukhov and A.~Semenov,
  Comput.\ Phys.\ Commun.\  {\bf 176}, 367 (2007)
  [hep-ph/0607059].

\bibitem{Vicente:2014wga}
  A.~Vicente and C.~E.~Yaguna,
  JHEP {\bf 1502}, 144 (2015)
  [arXiv:1412.2545 [hep-ph]].

\bibitem{Silveira:1985rk}
  V.~Silveira and A.~Zee,
  Phys.\ Lett.\ B {\bf 161}, 136 (1985).
  J.~McDonald,
  Phys.\ Rev.\ D {\bf 50}, 3637 (1994)
  [hep-ph/0702143 [HEP-PH]].
  C.~P.~Burgess, M.~Pospelov and T.~ter Veldhuis,
  Nucl.\ Phys.\ B {\bf 619}, 709 (2001)
  [hep-ph/0011335].
  V.~Barger, P.~Langacker, M.~McCaskey, M.~J.~Ramsey-Musolf and G.~Shaughnessy,
  Phys.\ Rev.\ D {\bf 77}, 035005 (2008)
  [arXiv:0706.4311 [hep-ph]].
  X.~G.~He, T.~Li, X.~Q.~Li, J.~Tandean and H.~C.~Tsai,
  Phys.\ Rev.\ D {\bf 79}, 023521 (2009)
  [arXiv:0811.0658 [hep-ph]].
  Y.~Mambrini,
  Phys.\ Rev.\ D {\bf 84}, 115017 (2011)
  [arXiv:1108.0671 [hep-ph]].
  M.~Gonderinger, H.~Lim and M.~J.~Ramsey-Musolf,
  Phys.\ Rev.\ D {\bf 86}, 043511 (2012)
  [arXiv:1202.1316 [hep-ph]].
  F.~S.~Queiroz, K.~Sinha and A.~Strumia,
  Phys.\ Rev.\ D {\bf 91}, 035006 (2015)
  [arXiv:1409.6301 [hep-ph]].
  H.~Han, J.~M.~Yang, Y.~Zhang and S.~Zheng,
  Phys.\ Lett.\ B {\bf 756}, 109 (2016)
  [arXiv:1601.06232 [hep-ph]].

\bibitem{Cline:2013gha}
  J.~M.~Cline, K.~Kainulainen, P.~Scott and C.~Weniger,
  Phys.\ Rev.\ D {\bf 88}, 055025 (2013)
  Erratum: [Phys.\ Rev.\ D {\bf 92}, 039906 (2015)]
  [arXiv:1306.4710 [hep-ph]].
  L.~Feng, S.~Profumo and L.~Ubaldi,
  JHEP {\bf 1503}, 045 (2015)
  [arXiv:1412.1105 [hep-ph]].
  M.~Duerr, P.~Fileviez Perez and J.~Smirnov,
  Phys.\ Lett.\ B {\bf 751}, 119 (2015)
  [arXiv:1508.04418 [hep-ph]].
  M.~Duerr, P.~Fileviez Perez and J.~Smirnov,
  arXiv:1509.04282 [hep-ph].
  A.~Beniwal, F.~Rajec, C.~Savage, P.~Scott, C.~Weniger, M.~White and A.~G.~Williams,
  arXiv:1512.06458 [hep-ph].

\bibitem{Aprile:2012zx}
  E.~Aprile [XENON1T Collaboration],
  Springer Proc.\ Phys.\  {\bf 148}, 93 (2013)
  [arXiv:1206.6288 [astro-ph.IM]];
  E.~Aprile {\it et al.} [XENON Collaboration],
  JCAP {\bf 1604}, 027 (2016)
  [arXiv:1512.07501 [physics.ins-det]].

\bibitem{Griest:1990kh}
  K.~Griest and D.~Seckel,
  Phys.\ Rev.\ D {\bf 43}, 3191 (1991).
  G.~Bertone, D.~Hooper and J.~Silk,
  Phys.\ Rept.\  {\bf 405}, 279 (2005)
  [hep-ph/0404175].

\bibitem{Ade:2013zuv}
  P.~A.~R.~Ade {\it et al.} [Planck Collaboration],
  Astron.\ Astrophys.\  {\bf 571}, A16 (2014)
  [arXiv:1303.5076 [astro-ph.CO]].

\bibitem{Akerib:2013tjd}
  D.~S.~Akerib {\it et al.} [LUX Collaboration],
  Phys.\ Rev.\ Lett.\  {\bf 112}, 091303 (2014)
  [arXiv:1310.8214 [astro-ph.CO]].
  D.~S.~Akerib {\it et al.} [LUX Collaboration],
  Phys.\ Rev.\ Lett.\  {\bf 116}, 161301 (2016)
  [arXiv:1512.03506 [astro-ph.CO]].

\bibitem{Akerib:2016vxi}
  D.~S.~Akerib {\it et al.} [LUX Collaboration],
  arXiv:1608.07648 [astro-ph.CO].

\bibitem{Tan:2016zwf}
  A.~Tan {\it et al.} [PandaX-II Collaboration],
  arXiv:1607.07400 [hep-ex].

\bibitem{Ackermann:2015zua}
  M.~Ackermann {\it et al.} [Fermi-LAT Collaboration],
  Phys.\ Rev.\ Lett.\  {\bf 115}, 231301 (2015)
  [arXiv:1503.02641 [astro-ph.HE]].

\bibitem{Ackermann:2015lka}
  M.~Ackermann {\it et al.} [Fermi-LAT Collaboration],
  Phys.\ Rev.\ D {\bf 91}, 122002 (2015)
  [arXiv:1506.00013 [astro-ph.HE]].

\bibitem{Abramowski:2013ax}
  A.~Abramowski {\it et al.} [HESS Collaboration],
  Phys.\ Rev.\ Lett.\  {\bf 110}, 041301 (2013)
  [arXiv:1301.1173 [astro-ph.HE]].

\bibitem{Khachatryan:2014jba}
  V.~Khachatryan {\it et al.} [CMS Collaboration],
  Eur.\ Phys.\ J.\ C {\bf 75}, 212 (2015)
  [arXiv:1412.8662 [hep-ex]].
  T.~Corbett, O.~J.~P.~Eboli, D.~Goncalves, J.~Gonzalez-Fraile, T.~Plehn and M.~Rauch,
  JHEP {\bf 1508}, 156 (2015)
  [arXiv:1505.05516 [hep-ph]].
  G.~Aad {\it et al.} [ATLAS Collaboration],
  JHEP {\bf 1511}, 206 (2015)
  [arXiv:1509.00672 [hep-ex]].
  G.~Aad {\it et al.} [ATLAS and CMS Collaborations],
  arXiv:1606.02266 [hep-ex].

\bibitem{Doro:2012xx}
  M.~Doro {\it et al.} [CTA Consortium Collaboration],
  Astropart.\ Phys.\  {\bf 43}, 189 (2013)
  [arXiv:1208.5356 [astro-ph.IM]].
  M.~Wood, J.~Buckley, S.~Digel, S.~Funk, D.~Nieto and M.~A.~Sanchez-Conde,
  arXiv:1305.0302 [astro-ph.HE].
  H.~Silverwood, C.~Weniger, P.~Scott and G.~Bertone,
  JCAP {\bf 1503}, 055 (2015)
  [arXiv:1408.4131 [astro-ph.HE]].
  V.~Lefranc, E.~Moulin, P.~Panci and J.~Silk,
  Phys.\ Rev.\ D {\bf 91}, 122003 (2015)
  [arXiv:1502.05064 [astro-ph.HE]].
  J.~Carr {\it et al.} [CTA Consortium Collaboration],
  arXiv:1508.06128 [astro-ph.HE].

\bibitem{Cirelli:2010xx}
  M.~Cirelli {\it et al.},
  JCAP {\bf 1103}, 051 (2011)
  Erratum: [JCAP {\bf 1210}, E01 (2012)]
  [arXiv:1012.4515 [hep-ph]].

\bibitem{Daylan:2014rsa}
  T.~Daylan, D.~P.~Finkbeiner, D.~Hooper, T.~Linden, S.~K.~N.~Portillo, N.~L.~Rodd and T.~R.~Slatyer,
  Phys.\ Dark Univ.\  {\bf 12}, 1 (2016)
  [arXiv:1402.6703 [astro-ph.HE]].
  T.~Mondal and T.~Basak,
  Phys.\ Lett.\ B {\bf 744}, 208 (2015)
  [arXiv:1405.4877 [hep-ph]].
  B.~Zhou, Y.~F.~Liang, X.~Huang, X.~Li, Y.~Z.~Fan, L.~Feng and J.~Chang,
  Phys.\ Rev.\ D {\bf 91}, 123010 (2015)
  [arXiv:1406.6948 [astro-ph.HE]].
  F.~Calore, I.~Cholis and C.~Weniger,
  JCAP {\bf 1503}, 038 (2015)
  [arXiv:1409.0042 [astro-ph.CO]].
  D.~Hooper and L.~Goodenough,
  Phys.\ Lett.\ B {\bf 697}, 412 (2011)
  [arXiv:1010.2752 [hep-ph]].
  L.~Goodenough and D.~Hooper,
  arXiv:0910.2998 [hep-ph].

\bibitem{TheFermi-LAT:2015kwa}
  M.~Ajello {\it et al.} [Fermi-LAT Collaboration],
  Astrophys.\ J.\  {\bf 819}, 44 (2016)
  [arXiv:1511.02938 [astro-ph.HE]].

\bibitem{GCE-astro}
  C.~Gordon and O.~Macias,
  Phys.\ Rev.\ D {\bf 88}, 083521 (2013)
  Erratum: [Phys.\ Rev.\ D {\bf 89}, 049901 (2014)]
  [arXiv:1306.5725 [astro-ph.HE]].
  K.~N.~Abazajian, N.~Canac, S.~Horiuchi and M.~Kaplinghat,
  Phys.\ Rev.\ D {\bf 90}, 023526 (2014)
  [arXiv:1402.4090 [astro-ph.HE]].
  J.~Petrovic, P.~D.~Serpico and G.~Zaharijas,
  JCAP {\bf 1410}, 052 (2014)
  [arXiv:1405.7928 [astro-ph.HE]].
  R.~Bartels, S.~Krishnamurthy and C.~Weniger,
  Phys.\ Rev.\ Lett.\  {\bf 116}, 051102 (2016)
  [arXiv:1506.05104 [astro-ph.HE]].
  I.~Cholis, C.~Evoli, F.~Calore, T.~Linden, C.~Weniger and D.~Hooper,
  JCAP {\bf 1512}, 005 (2015)
  [arXiv:1506.05119 [astro-ph.HE]].
  S.~K.~Lee, M.~Lisanti, B.~R.~Safdi, T.~R.~Slatyer and W.~Xue,
  Phys.\ Rev.\ Lett.\  {\bf 116}, 051103 (2016)
  [arXiv:1506.05124 [astro-ph.HE]].

\bibitem{Hooper:2011ti}
  D.~Hooper and T.~Linden,
  Phys.\ Rev.\ D {\bf 84}, 123005 (2011)
  [arXiv:1110.0006 [astro-ph.HE]].
  D.~Hooper and T.~R.~Slatyer,
  Phys.\ Dark Univ.\  {\bf 2}, 118 (2013)
  [arXiv:1302.6589 [astro-ph.HE]].
  P.~Agrawal, B.~Batell, P.~J.~Fox and R.~Harnik,
  JCAP {\bf 1505}, 011 (2015)
  [arXiv:1411.2592 [hep-ph]].
  F.~Calore, I.~Cholis, C.~McCabe and C.~Weniger,
  Phys.\ Rev.\ D {\bf 91}, 063003 (2015)
  [arXiv:1411.4647 [hep-ph]].

\bibitem{Cuoco:2016jqt}
  A.~Cuoco, B.~Eiteneuer, J.~Heisig and M.~Kramer,
  arXiv:1603.08228 [hep-ph].
  F.~S.~Sage and R.~Dick,
  arXiv:1604.04589 [astro-ph.HE].
  M.~Duerr, P.~Fileviez Perez and J.~Smirnov,
  arXiv:1510.07562 [hep-ph].
  J.~D.~Ruiz-Alvarez, C.~A.~de S.Pires, F.~S.~Queiroz, D.~Restrepo and P.~S.~Rodrigues da Silva,
  Phys.\ Rev.\ D {\bf 86}, 075011 (2012)
  [arXiv:1206.5779 [hep-ph]].
  
\bibitem{Balazs:2014jla}
  C.~Balazs and T.~Li,
  Phys.\ Rev.\ D {\bf 90}, 055026 (2014)
  [arXiv:1407.0174 [hep-ph]].
  M.~Kaplinghat, T.~Linden and H.~B.~Yu,
  Phys.\ Rev.\ Lett.\  {\bf 114}, 211303 (2015)
  [arXiv:1501.03507 [hep-ph]].
  P.~Ko and Y.~Tang,
  JCAP {\bf 1602}, 011 (2016)
  [arXiv:1504.03908 [hep-ph]].
  C.~Balazs, T.~Li, C.~Savage and M.~White,
  Phys.\ Rev.\ D {\bf 92}, 123520 (2015)
  [arXiv:1505.06758 [hep-ph]].
  A.~Biswas, S.~Choubey and S.~Khan,
  arXiv:1604.06566 [hep-ph].
  N.~Okada and O.~Seto,
  Phys.\ Rev.\ D {\bf 89}, 043525 (2014)
  [arXiv:1310.5991 [hep-ph]].

\bibitem{Craig:2014lda}
  N.~Craig, H.~K.~Lou, M.~McCullough and A.~Thalapillil,
  JHEP {\bf 1602}, 127 (2016)
  [arXiv:1412.0258 [hep-ph]].

\bibitem{Aad:2015txa}
  G.~Aad {\it et al.} [ATLAS Collaboration],
  JHEP {\bf 1601}, 172 (2016)
  [arXiv:1508.07869 [hep-ex]].

\bibitem{Chatrchyan:2014tja}
  S.~Chatrchyan {\it et al.} [CMS Collaboration],
  Eur.\ Phys.\ J.\ C {\bf 74}, 2980 (2014)
  [arXiv:1404.1344 [hep-ex]].

\bibitem{Aad:2014iia}
  G.~Aad {\it et al.} [ATLAS Collaboration],
  Phys.\ Rev.\ Lett.\  {\bf 112}, 201802 (2014)
  [arXiv:1402.3244 [hep-ex]].

\bibitem{Bernaciak:2014pna}
  C.~Bernaciak, T.~Plehn, P.~Schichtel and J.~Tattersall,
  Phys.\ Rev.\ D {\bf 91}, 035024 (2015)
  [arXiv:1411.7699 [hep-ph]].

\bibitem{Djouadi:2005gi}
  A.~Djouadi,
  Phys.\ Rept.\  {\bf 457}, 1 (2008)
  [hep-ph/0503172].
  S.~Heinemeyer {\it et al.} [LHC Higgs Cross Section Working Group Collaboration],
  arXiv:1307.1347 [hep-ph].

\bibitem{Baek:2014jga}
  S.~Baek, P.~Ko and W.~I.~Park,
  Phys.\ Rev.\ D {\bf 90}, 055014 (2014)
  [arXiv:1405.3530 [hep-ph]].

\bibitem{Degrande:2011ua}
  C.~Degrande, C.~Duhr, B.~Fuks, D.~Grellscheid, O.~Mattelaer and T.~Reiter,
  Comput.\ Phys.\ Commun.\  {\bf 183}, 1201 (2012)
  [arXiv:1108.2040 [hep-ph]].

\bibitem{Alwall:2011uj}
  J.~Alwall, M.~Herquet, F.~Maltoni, O.~Mattelaer and T.~Stelzer,
  JHEP {\bf 1106}, 128 (2011)
  [arXiv:1106.0522 [hep-ph]].
  J.~Alwall {\it et al.},
  JHEP {\bf 1407}, 079 (2014)
  [arXiv:1405.0301 [hep-ph]].

\bibitem{Ball:2012cx}
  R.~D.~Ball {\it et al.},
  Nucl.\ Phys.\ B {\bf 867}, 244 (2013)
  [arXiv:1207.1303 [hep-ph]].
  R.~D.~Ball {\it et al.} [NNPDF Collaboration],
  JHEP {\bf 1504}, 040 (2015)
  [arXiv:1410.8849 [hep-ph]].

\bibitem{Sjostrand:2006za}
  T.~Sjostrand, S.~Mrenna and P.~Z.~Skands,
  JHEP {\bf 0605}, 026 (2006)
  [hep-ph/0603175].

\bibitem{Ovyn:2009tx}
  S.~Ovyn, X.~Rouby and V.~Lemaitre,
  arXiv:0903.2225 [hep-ph].
  J.~de Favereau {\it et al.} [DELPHES 3 Collaboration],
  JHEP {\bf 1402}, 057 (2014)
  [arXiv:1307.6346 [hep-ex]].

\bibitem{Conte:2012fm}
  E.~Conte, B.~Fuks and G.~Serret,
  Comput.\ Phys.\ Commun.\  {\bf 184}, 222 (2013)
  [arXiv:1206.1599 [hep-ph]].

\bibitem{Chatrchyan:2012jua}
  S.~Chatrchyan {\it et al.} [CMS Collaboration],
  JINST {\bf 8}, P04013 (2013)
  [arXiv:1211.4462 [hep-ex]].
  G.~Aad {\it et al.} [ATLAS Collaboration],
  JINST {\bf 11}, P04008 (2016)
  [arXiv:1512.01094 [hep-ex]].

\bibitem{Chatrchyan:2014aea}
  S.~Chatrchyan {\it et al.} [CMS Collaboration],
  Phys.\ Rev.\ D {\bf 90}, 032006 (2014)
  [arXiv:1404.5801 [hep-ex]].

\bibitem{Aad:2014iza}
  G.~Aad {\it et al.} [ATLAS Collaboration],
  Phys.\ Rev.\ D {\bf 90}, 052001 (2014)
  [arXiv:1405.5086 [hep-ex]].
\bibitem{Aad:2014nua}
  G.~Aad {\it et al.} [ATLAS Collaboration],
  JHEP {\bf 1404}, 169 (2014)
  [arXiv:1402.7029 [hep-ex]].

\bibitem{Aad:2015eda}
  G.~Aad {\it et al.} [ATLAS Collaboration],
  Phys.\ Rev.\ D {\bf 93}, 052002 (2016)
  [arXiv:1509.07152 [hep-ex]].

\end{thebibliography}
\end{document}